\documentclass{aa}

\usepackage{graphicx}
\usepackage{txfonts}
\usepackage{booktabs}
\usepackage[tableposition=top]{caption}

\usepackage{hyperref}

\usepackage{xcolor}

\usepackage{placeins}

\usepackage{bm}
\usepackage{amsmath}

\usepackage{dblfloatfix}

\newcommand{\lya}{Lyman-$\alpha$}
\newcommand{\lyb}{Lyman-$\beta$}
\newcommand{\Nv}{\ion{N}{v}}
\newcommand{\Civ}{\ion{C}{iv}}
\newcommand{\MgII}{\ion{Mg}{ii}}
\newcommand{\Hb}{H-$\beta$}
\newcommand{\FeII}{\ion{Fe}{ii}}
\newcommand{\OVI}{\ion{O}{vi}}

\newcommand{\angstrom}{\textup{\AA}}

\begin{document}
   \title{QUEST: Quasar Unsupervised Encoder and Synthesis Tool}

   \subtitle{A machine learning framework to generate quasar spectra}

   \author{F. Guarneri \inst{1, 2}\thanks{francesco.guarneri@uni-hamburg.de -- \url{https://github.com/cosmic-dawn-group/QUEST}},
           J. T. Schindler \inst{1},
           R. A. Meyer \inst{3},
           D. Yang \inst{4},
           J. F. Hennawi \inst{4, 5},
           L. Lucie-Smith \inst{1},
           S. E. I. Bosman \inst{6, 7},
           F. B. Davies \inst{7}
           }

    \institute{Hamburger Sternwarte, Universit\"at Hamburg, Gojenbergsweg 112, D-21029 Hamburg, Germany
    \and
    INAF--Osservatorio Astronomico di Trieste, Via G.B. Tiepolo, 11, I-34143 Trieste, Italy
    \and
    Department of Astronomy, University of Geneva, Chemin Pegasi 51, 1290 Versoix, Switzerland
    \and
    Leiden Observatory, Leiden University, P.O. Box 9513, 2300 RA Leiden, The Netherlands
    \and
    Department of Physics, University of California, Santa Barbara, CA 93106, USA
    \and
    Institute for Theoretical Physics, Heidelberg University, Philosophenweg 12, D–69120, Heidelberg, Germany
    \and
    Max-Planck-Institut f\"{u}r Astronomie, K\"{o}nigstuhl 17, 69117 Heidelberg, Germany
    }

   \date{Received ...; accepted ...}

  \abstract
   {Quasars at the redshift frontier ($z > 7.0$) are fundamental probes of black hole growth and evolution but notoriously difficult to identify. At these redshifts, machine learning-based selection methods have proven to be efficient, but require appropriate training sets to express their full potential.}
   {Here, we present \texttt{QUEST}, a Variational Auto-Encoder capable of generating realistic quasar spectra that can be post-processed for the generation of synthetic photometry and for spectral imputation.}
   {We start from the SDSS DR16Q catalogue, pre-process the spectra, and vet the sample to obtain a clean data set. After training the model, we investigate the properties of its latent space to understand whether it has learnt relevant physics.
   Furthermore, we provide a pipeline to generate photometry from the sampled spectra, compare it with actual quasar photometry, and showcase the capabilities of the model in reconstructing and extending quasar spectra.}
   {The trained network faithfully reproduces the input spectrum, both in terms of sample median and variance. By examining the latent space, we find correlations with, among others, continuum and bolometric luminosity, black hole mass, redshift, continuum slope, and emission line properties. When used to generate photometry, we find results in excellent agreement with the control sample. The model provides satisfactory results in reconstructing emission lines: estimates of the black hole mass from the reconstructed spectra are in good agreement with those from the original SDSS spectra. Furthermore, when spectra with broad absorption line features are reconstructed, the model successfully interpolates over the absorption systems. Compared with previous work, we find excellent agreement between the spectra sampled from our model and the output of their results. However, \texttt{QUEST} does not require any ad-hoc tuning, and is capable of reproducing the full variety of spectra available in the training set.}
   {}


   \authorrunning{F. Guarneri et al.}
   \maketitle

\section{Introduction}
Quasars, actively accreting supermassive black holes (SMBH), are the most luminous Active Galactic Nuclei (AGNs) and non-transient sources in the sky \citep[see][for a recent review]{fan_quasars_2023}. Their luminosity (typically $\log{\left(L_{\rm BOL}\right)}~\sim~46-48$ erg s$^{-1}$) makes them detectable out to redshift $z > 7.5$, when the Universe was less than a Gyr old \citep{banados_800-million-solar-mass_2018, yang_poniuaena_2020, wang_luminous_2021}. Their existence places stringent constraints on the growth history and seeding mechanisms of SMBH \citep[e.g.][]{yang_probing_2021}. Their physical distance allows us to investigate the epoch of reionisation \citep[e.g., ][]{kist_quantifying_2025} and the chemical and physical state of the intergalactic medium \citep[IGM, e.g.][]{wang_significantly_2020}. Quasars shape and influence their surrounding environment: molecular outflows have been detected in samples of quasars at $z > 6.0$ \citep{spilker_direct_2025}, and feedback from these objects is often invoked to quench galaxies. Quasars themselves are often found to live in overdense regions \citep{meyer_constraining_2022, wang_spectroscopic_2023, champagne_mixture_2023} and are hosted by the most massive \citep{neeleman_kinematics_2021} and star-forming \citep{salvestrini_molecular_2025} galaxies in the Universe. Compared to local AGNs, quasars are often found to be overmassive in relation to their host galaxy (e.g. \citealt{farina_x-shooteralma_2022}, but see also \citealt{li_connection_2022, silverman_shellqs-jwst_2025}). However, despite decades of quasar investigations, many of these topics remain open questions: it is unclear what the main seeding and evolution pathways are to grow these objects in such a short amount of time \citep[see, for example,][for reviews on the topic]{inayoshi_assembly_2020, volonteri_origins_2021}. %
A precise timeline for reionisation still eludes us \citep{durovcikova_chronicling_2024, qin_percent-level_2025, umeda_probing_2025}.
Although it is well established that quasars and their host galaxy co-evolve \citep{kormendy_coevolution_2013}, at $z \gtrsim 6.0$ the results of clustering analyses point towards a very diverse environment (see, for example, \citealt{meyer_constraining_2022, champagne_mixture_2023} and references therein).

In order to effectively investigate these problems, large and well-defined samples of quasars at different redshifts are needed. In the last 20 years, in particular, a lot of effort has been devoted to pushing the quasar redshift frontier
further into the epoch of reionisation, from redshift $z\approx6$ \citep{fan_constraining_2006} to $z\approx 7.64$ \citep{wang_luminous_2021}, to characterise the high-$z$ quasar population \citep{wang_luminous_2021}. Sensitive near-infrared surveys over wide areas and careful quasar selection techniques were critical to this success. Several methods have been applied in the search for quasars, but most of the known high-redshift quasar population has been identified through standard colour selections \citep{banados_pan-starrs1_2023, belladitta_discovery_2025}. However, at $z \gtrsim 7.5$, these methods are only about 1\% efficient \citep{nanni_paving_2022}, 
making large spectroscopic follow-up campaigns unfeasible for future space-based surveys (such as \textit{Euclid}, \citealt{euclid_collaboration_euclid_2025}, or the Nancy Grace Roman Space Telescope) that will yield an order of magnitude more sources than ground-based counterparts.

In preparation for these surveys, statistical methods have been developed and applied with excellent results. Bayesian selection algorithms had already yielded the first quasar at $z > 7$ \citep{mortlock_luminous_2011} and were successfully applied to define complete samples in the VIKING footprint \citep{edge_vista_2013, barnett_complete_2021}. Although effective, these methods depend on prior assumptions about contaminant populations, which are poorly constrained at faint magnitudes. More recently, machine learning (ML) techniques have been employed, providing probabilistic classifications of objects and much higher selection efficiencies \citep[$\geq 15$\%, ][]{wenzl_random_2021, nanni_paving_2022, yang_high-z_2024, kang_extreme_2024, byrne_quasar_2024}. Nevertheless, a significant limitation still exists: the paucity of training data. At present, only 11 \citep[including the latest discovery in][]{matsuoka_subaru_2025} quasars with $z \geq 7.0$ are known. This scarcity can be alleviated through the generation of synthetic datasets: generative machine learning models are perfectly suited for this task.

In this paper, we present an Information Maximising Variational Auto-Encoder (Info-VAE) trained to produce realistic quasar spectra. These can be post-processed to generate reliable and accurate photometry, which will be used in turn to identify the highest redshift quasars in upcoming photometric surveys (such as the \textit{Euclid} Wide Survey, the Legacy Survey of Space and Time, and the Roman Wide Survey). Such a model can be naturally extended to several other applications: \textit{i}) reconstruct the quasar continuum in the \lya{} forest \textit{ii}) extend quasar spectra to bluer or redder wavelengths; \textit{iii}) reconstruct regions affected by telluric lines or by broad absorption lines (BAL); \textit{iv}) ambitiously, reconstruct emission lines, in order to estimate the quasar black hole mass through single epoch virial estimators \citep[an approach complementary to previous works, such as][where the BH mass is directly estimated]{eilers_generative_2022}.

The paper is organised as follows: Section \ref{sec:training_dataset} details the approach to generate the training datasets and summarises the most relevant information. Section \ref{sec:VAE_design} provides a general introduction to VAEs, describes our implementation, and outlines our hyperparameter optimisation strategy. In Section \ref{sec:latent_space_exploration}, we examine the properties of the latent space and assess whether the model has learnt relevant quasar physics. Section \ref{sec:VAE_application} presents various use cases for the Info-VAE, demonstrating its capabilities. We compare our results with previous studies and discuss the known limitations in Section \ref{sec:discussion}, and conclude in Section \ref{sec:conclusion}. We adopt the following cosmological parameters: $\Omega_{\rm m}=0.3111$, $\Omega_\Lambda=0.6899$, $h=0.6766$ \citep{planck_collaboration_planck_2020}. All magnitudes are presented in the AB system \citep{oke_secondary_1983}.

\section{Training datasets}
\label{sec:training_dataset}
In this section, we will describe the datasets used to train the Info-VAEs. We assembled three datasets and used them to train three different models with the same network architecture. All datasets are processed uniformly and differ only in terms of the minimum signal-to-noise ratio and wavelength coverage required. Each dataset is used to train the corresponding VAE model (see Sect. \ref{sec:optimisation}). In the following, we will refer to these datasets and the corresponding models as ``General Purpose'' (GP)'', ``Full Overlap Blue'' (FOB) and ``Full Overlap Red'' (FOR) datasets. 

\subsection{The ``General Purpose'' dataset}
\label{sect:GP_dataset_prep}
We aim to train a machine learning model capable of several tasks: \textit{i)} generating realistic quasar spectra and photometry \textit{ii)} inputting a quasar spectrum to regions that were not originally covered by the SDSS spectrograph; \textit{iii)} reconstructing intermediate regions of the quasar spectrum that are contaminated by BALs or affected by instrumental and observational systematics; \textit{iv)} faithfully reconstructing selected emission lines and allow, for example, computation of the quasar's black hole mass. 

\begin{figure}[ht]
    \centering
    \includegraphics[width=\columnwidth]{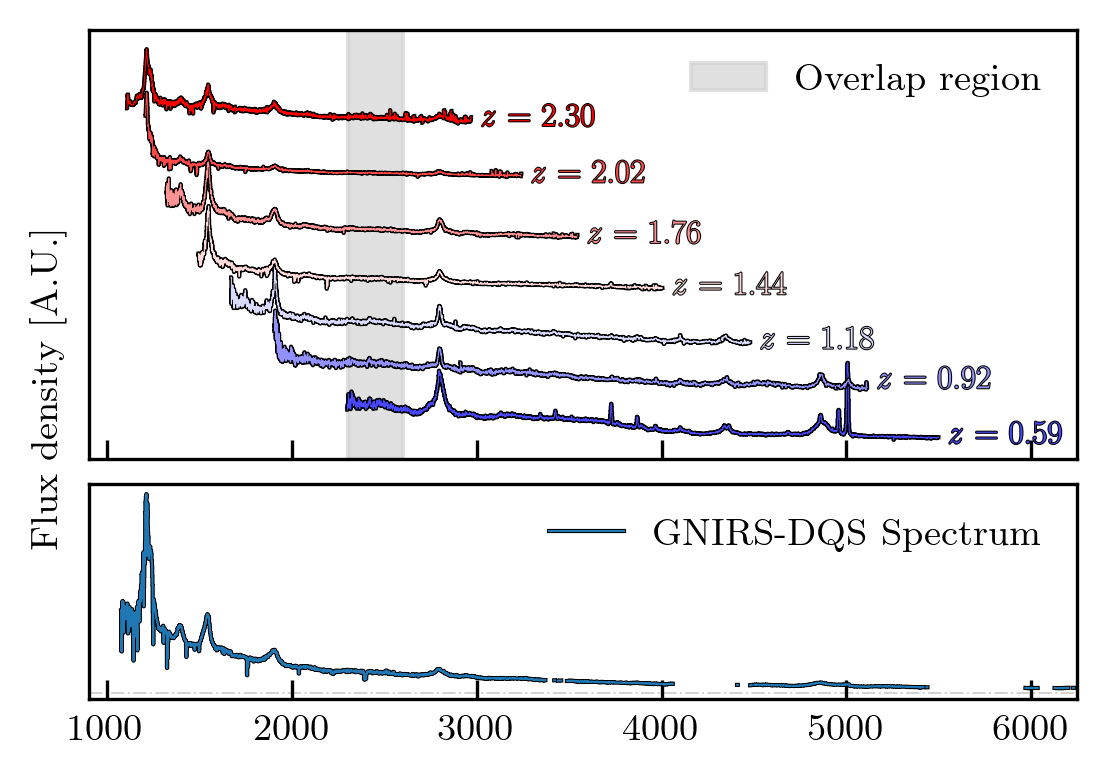}
    \caption{Top panel: Example of SDSS spectra at different redshifts included in the GP dataset. Bottom panel: Example of spectrum from the GNIRS-DQS survey, highlighting the extended coverage at redder wavelengths. Spectral gaps are are due to telluric absorptions.}
    \label{fig:SDSS_example_good}
\end{figure}

To be able to perform these tasks, we require a model capable of generating spectra that cover a large (rest-frame) wavelength range. For the purpose of this paper, this is chosen to be between 980~\AA{} and 5500~\AA{}, to cover the entirety of the \lyb{} forest, and UV and optical emission lines up to the H$\beta$--\ion{O}{[iii]} complex. Unfortunately, no large spectroscopic survey provides data that fully covers this range. However, it is possible to assemble a dataset in which spectra at different redshifts contribute to different portions of this wavelength space. In the case of the SDSS, for example, low-$z$ spectra cover the reddest wavelengths we require, while higher-$z$ spectra cover the bluest ones (Fig. \ref{fig:SDSS_example_good}).

\begin{figure}[ht]
    \centering
    \includegraphics[width=\columnwidth]{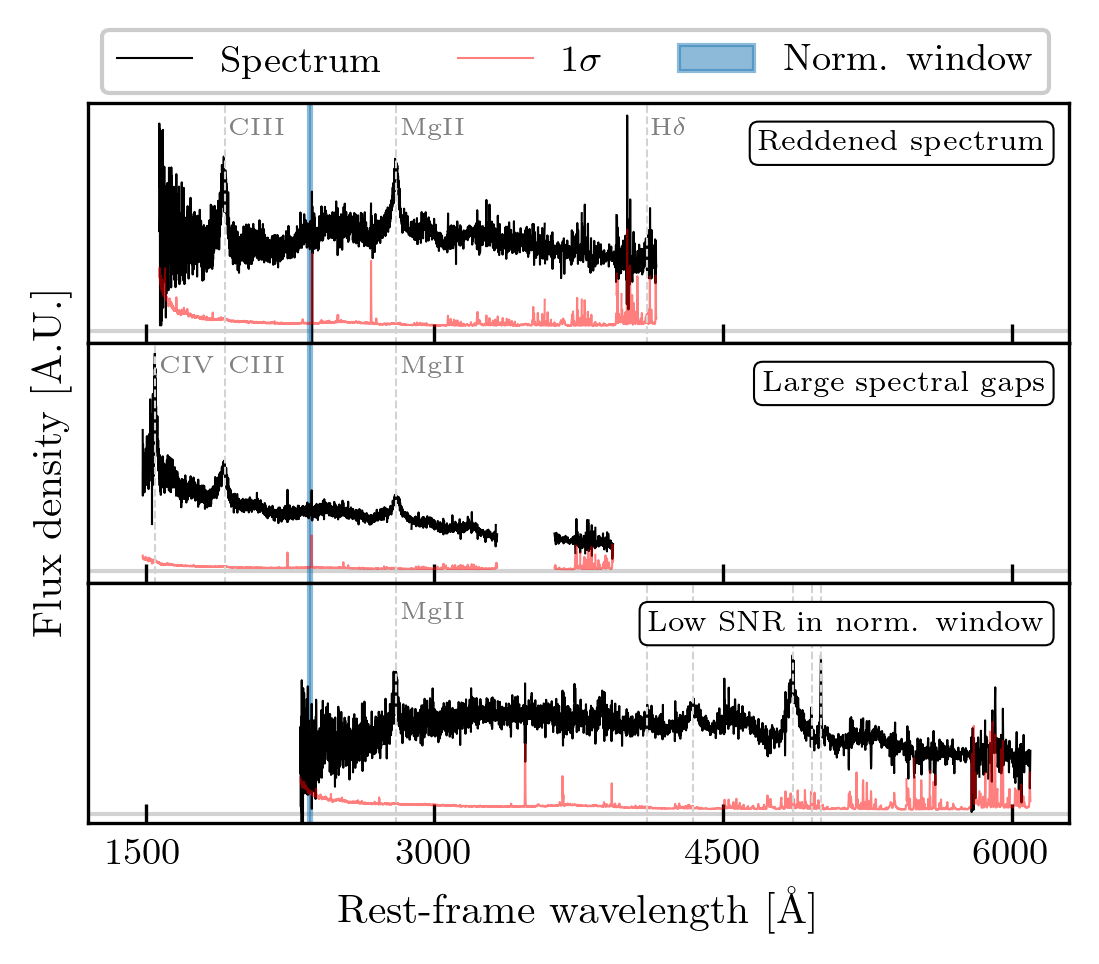}
    \caption{Examples of rejected spectra, and the cause of rejection. The black line shows the original SDSS spectrum, the red one the nominal uncertainty and the light grey one the zero-flux level.}
    \label{fig:SDSS_rejected_example}
\end{figure}

In addition to this, it is possible to combine spectra of the same object, collected with different facilities, to cover a larger wavelength range. In order to assemble an optimal training set, we start from the SDSS DR16 quasar catalogue \citep[][hereafter DR16Q]{lyke_sloan_2020}. We opt for a mature and well-studied sample, rather than more recent ones \citep[such as, for instance, the first data release from the DESI Collaboration][]{desi_collaboration_data_2025}. This allows us to exploit ancillary data made available by the community \citep[e. g.][]{2022ApJS..263...42W}, and complement and extend SDSS spectra to redder wavelengths including publicly available near-infrared data from the GNIRS-DQS survey \citep{matthews_placing_2021, matthews_gemini_2023}. 
Spectra were independently re-reduced: additional details will be presented in a forthcoming publication (Yang et al., in prep.).
As a first step, we collect all spectra that satisfy simple quality cuts\footnote{We indicate in typeface font the column names of the SDSS DR16Q catalogue.}:
\begin{itemize}
    \item 0.59 < \texttt{Z\_PIPE} < 2.77 and \texttt{ZWARNING} = 0, to select reliable redshifts and guarantee that, taken together, the spectra fully cover the aforementioned wavelength range while allowing a common overlap region between 2300--2600~\AA{}. Although the wavelength range over which we require the overlap is arbitrary, it is important to include at least an emission line-free region to consistently normalise all inputs. We choose the normalisation region to be between 2350~\AA{} and 2360~\AA{} (rest-frame wavelength);
    \item \texttt{BI\_CIV} $\leq 0$ and \texttt{BI\_SiIV} $\leq 0$, to remove quasars with the most prominent BAL features;
    \item \texttt{SN\_MEDIAN\_ALL}\footnote{Defined, according to the SDSS documentation, as ``Median S/N value of all good spectroscopic pixels.''} > 15 and \texttt{M\_I} < -20, to only include spectra of bright quasars with sufficient S/N to clearly detect continuum, emission lines and weaker BAL features.
\end{itemize}
This results in a parent sample that contains 20~007 quasars, with a median redshift of 1.62 and absolute \textit{i}-band magnitude -26.82. Once collected, all the spectra that satisfy these simple cuts are further preprocessed and analysed to discard those with artefacts, large interpolated regions, and weaker BALs features that were not excluded by the balinicity index cut previously imposed. In particular, we further clean up the sample by identifying and excluding spectra with at least fifteen consecutive interpolated pixels, without any flux density value in the normalisation window, or for which the median S/N in the normalisation window is lower than seven. Furthermore, we exclude reddened spectra, spectra with broad absorption features on the blue or red side of the \lya{} and \Civ{} emission lines via a custom automated pipeline, and spectra with fewer than 100 valid pixels. We show examples of rejected spectra in Fig. \ref{fig:SDSS_rejected_example}.
The cleaning procedure excludes 1786 spectra, leaving us with a data set of 18~221 objects that we deem usable for training. For all these spectra, we:
\begin{itemize}
    \item shift them to the respective rest frame, by dividing the wavelength axis and multiplying the quasar flux density by $(1+z_{\rm quasar})$. We use \texttt{Z\_PIPE} as the fiducial quasar redshift;
    \item correct for the effect of the Milky Way's dust extinction by de-reddening each spectrum using the \citet{2023ApJ...950...86G} $A(\lambda)$ extinction curve. We assume an average value $R(V) = 3.1$, as is commonly done for the Milky Way \citep[see e.g.][]{1980MNRAS.192..467W, 1999ApJ...525.1011F} and compute the $E(B-V)$ at the quasar coordinates $(l, b)_{\rm quasars}$ based on the two dimensional dust map from \citet{2023ApJ...958..118C}
    \item fit a continuum to the quasar spectrum, following an approach similar to that adopted in \citet{bosman_comparison_2021}, developed by \citet{1979ApJ...229..891Y, 1982MNRAS.198...91C} and first implemented in \citet{2008A&A...491..465D}. Briefly, the algorithm fits a spline over equally spaced nodes along the quasar spectrum. During the fitting, individual pixels are iteratively masked via asymmetric sigma-clipping. Iterations are stopped, and convergence is reached when the standard deviation of the fluxes in the retained pixels is less than the average observed noise. The fitted continuum will be used to further clean up the sample to remove weak BALs and replace the Lyman forest of the spectrum with unabsorbed flux (see below).
    \item normalising each spectrum by dividing the flux density by the median flux density in a wavelength region between 2350--2360~\AA{}. 
\end{itemize}
As a final step, we resample all the spectra on a common wavelength grid, from 980~\AA{} to 5500~\AA{}, linearly spaced in velocity space. We set the pixel size to 140~{km s$^{-1}$}. For all high-$z$ spectra we replace the \lya{} forest with the fitted continuum, smoothly joining the latter with the original spectrum around 1225~\AA{}. This step is necessary: for instance, to generate synthetic photometry of quasars with $z \gtrsim 2.0$, the suppression of the flux blueward of the \lya{} due to the intergalactic medium should be computed on the basis of the unabsorbed continuum.

\begin{figure}[ht]
    \centering
    \includegraphics[width=\columnwidth]{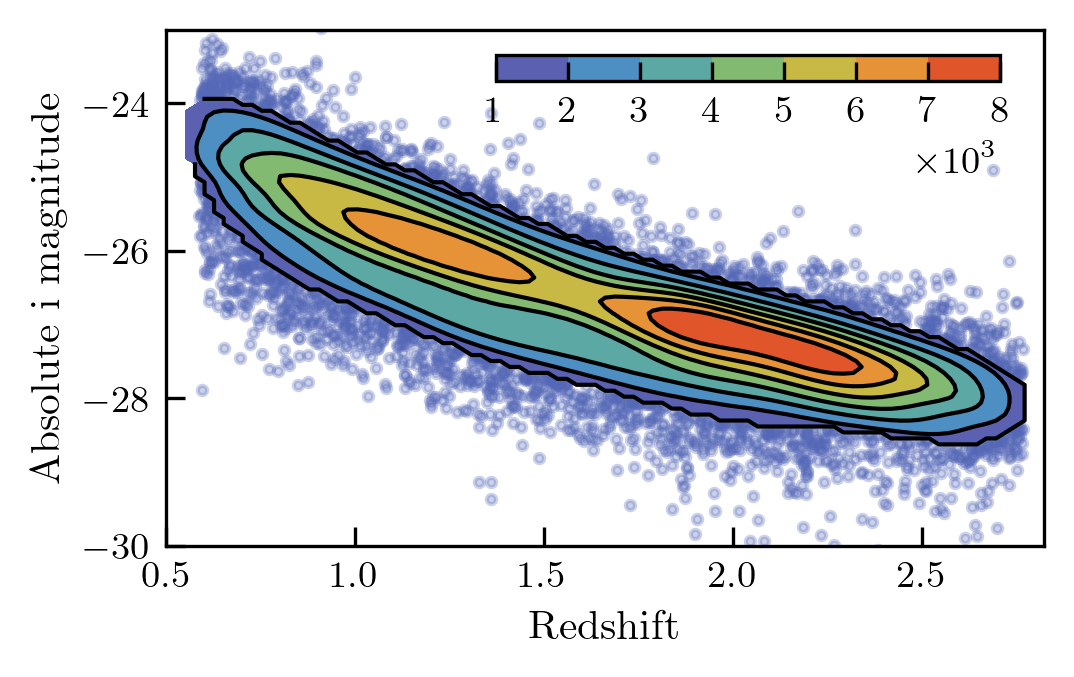}
    \caption{Density plot showing the redshift-absolute \textit{i}-band magnitude distribution for the spectra that meet the selection criteria. The colour map shows the number of spectra in each contour line.}
    \label{fig:MI-z_dist}
\end{figure}

\begin{figure}[ht]
    \centering
    \includegraphics[width=\columnwidth]{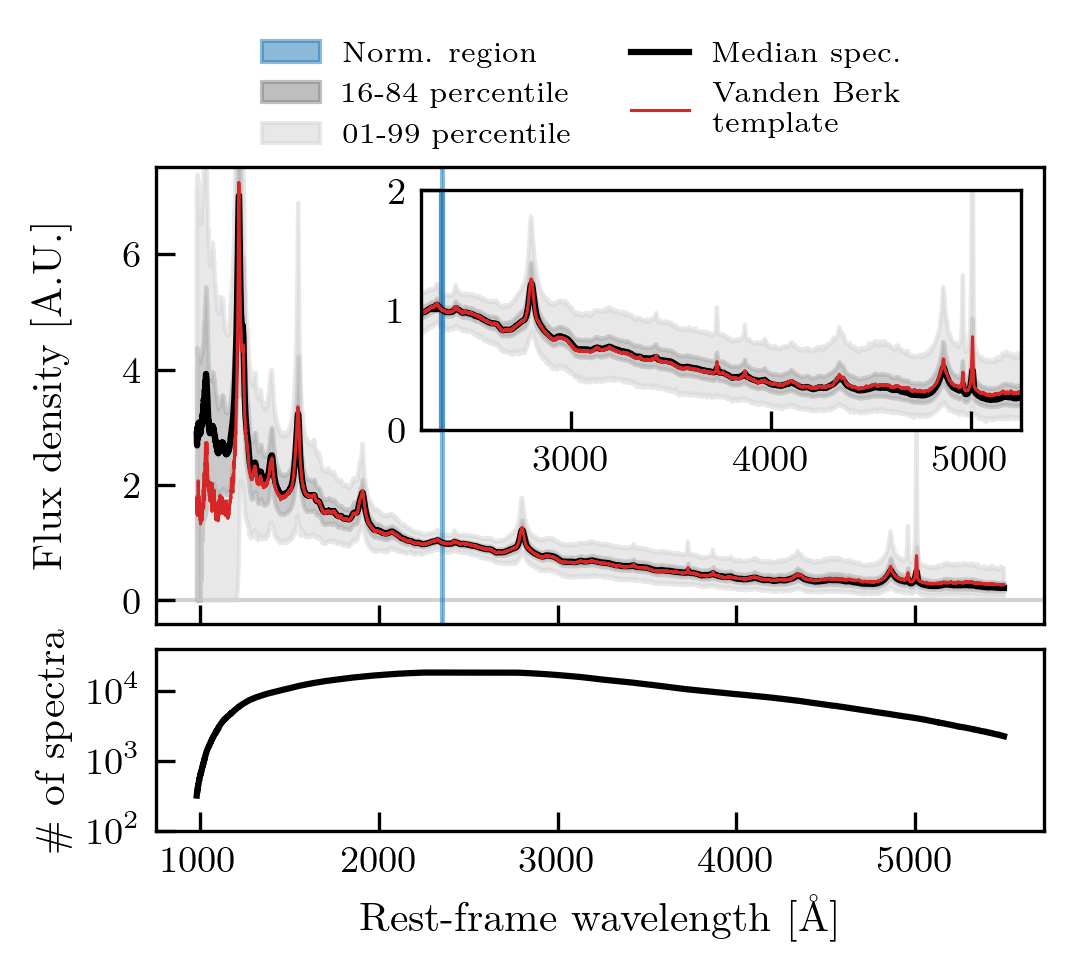}
    \caption{Median spectrum and logarithmic number of spectra contributing to the median in each pixel. Top panel: median spectrum of the quasars included in the training set (black, thick, solid line) compared to the \citet{vanden_berk_composite_2001} (red, thin, solid line). Shaded regions represent the 16$^{\rm th}$--84$^{\rm th}$ percentiles (dark grey) and the 1$^{\rm st}$--99$^{\rm th}$ percentile (light grey). The vertical, blue band represents the region in which we compute the normalisation factor for each spectrum. Bottom panel: logarithmic number of spectra contributing to each pixel in the median spectrum. By construction, all spectra contribute to the region between 2300--2600 \AA{}. 
    }
    \label{fig:medianCompositeSpec}
\end{figure}

The sample used for training has a median redshift of 1.61 (16$^{\rm th}$--84$^{\rm th}$ percentiles: 0.95--2.27, respectively) and median absolute \textit{i}-band magnitude of $-26.79$, (16$^{\rm th}$--84$^{\rm th}$ percentiles: -27.73 -- -25.47, respectively). We show a density plot with the distribution in the $z$--M$_{\rm i}$ plane in Fig. \ref{fig:MI-z_dist}, and a composite spectrum of all quasars that meet the selection criteria in Fig. \ref{fig:medianCompositeSpec}. 
We compare the latter with the type-1 quasar template from \citet[solid, red line in Fig. \ref{fig:medianCompositeSpec}]{vanden_berk_composite_2001} and provide the complete median composite in Table \ref{tab:medianCompositeSpec}. The median spectrum and the template agree very well redward of the \lya{} emission line; the continuum in the \lya{} forest is instead higher in our median spectrum, as a consequence of the grafting of the fitted continuum (see below). 

\subsection{The ``Full Overlap Blue'' and ``Full Overlap Red'' datasets}
We follow the same approach as outlined in the previous section to prepare these datasets. In both cases, we start from the SDSS DR16Q, select quasars that meet basic quality cuts, and process the sample. We lower the S/N threshold to 5 and require full coverage of the wavelengths 1175~\AA{}--2950~\AA{} (FOB) and 2300~\AA{}--5500~\AA{} (FOR).
Although suboptimal, lowering the S/N threshold is needed because of the lower number of spectra available, which turned out to be insufficient to train the VAE. We summarise the most important information for the three datasets in Table \ref{tab:summary_dataset}, including the total number of sources, the overlap range we require and the median redshift and absolute $i$-band magnitude.

\begin{table}[t]
    \centering
    \caption{Summary of the most relevant information for each dataset. We list the total number of sources left after the full cleaning process, the overlap range we require and the median redshift and absolute M$_{\rm i}$ as given by the SDSS DR16Q.}
    \begin{tabular}{c|c|c|c|c}
    \toprule
    Dataset & \#sources & Overlap range [\AA] & $z$ & M$_{\rm i}$ \\
    \midrule
    GP      & 18~221 & 2300--2600 & 1.61 & -26.79 \\
    FOB     & 14~563 & 1175--2950 & 2.21 & -26.36 \\
    FOR     & 12~568 & 2300--5500 & 0.70 & -23.66 \\
    \bottomrule
    \end{tabular}
    \label{tab:summary_dataset}
\end{table}

\section{Design of the Info-VAE for QSO spectra}
\label{sec:VAE_design}
Variational Auto-Encoders \citep[VAEs,][]{kingma_auto-encoding_2013} are unsupervised generative networks that map, in a probabilistic manner, high-dimensional data to a lower-dimensional representation. This low-dimensional representation, with dimension $\mathcal{D}$, is generally referred to as a latent space and, by design, should reflect the most meaningful properties of the data.

From an architecture point of view, a VAE is similar to a standard auto-encoder \citep[AE,][]{rumelhart_learning_1987} and consists of two networks chained together: an encoder that compresses the data and performs (non)linear dimensionality reduction, and a decoder that takes samples from the latent space distributions and reconstructs them to the higher-dimensional input representation. The key difference from an AE is in the interpretation of the latent representation $\mathbf{z}$ of a given input $\mathbf{x}$: in a VAE, this is a probability distribution function $p(\mathbf{z}|\mathbf{x})$; in an AE, it is instead a single point. In principle, this distribution could assume any form. In practice, however, it is generally assumed to be a multivariate Gaussian, that is $p(\mathbf{z}|\mathbf{x}) \sim \mathcal{N}(\boldsymbol{\mu}, \boldsymbol{\sigma})$, where $\boldsymbol{\mu}$ and $\boldsymbol{\sigma}$ are the output of the encoder and represent the means and standard deviations of the Gaussian distributions describing each latent space dimension. $\boldsymbol{\mu}$ and $\boldsymbol{\sigma}$ are the key ingredients in building the latent space dimensions $z_i$ using the reparametrisation trick: $\boldsymbol{z} = \boldsymbol{\mu} + \epsilon \boldsymbol{\sigma}$, with $\epsilon \sim \mathcal{N}(0, 1)$. Finally, the decoder takes the latent space as input and returns a distribution of reconstructed outputs $\mathbf{x'}$.

In order to train the algorithm, one needs to define an objective function to minimise. In the standard VAE implementation, this is taken to be the evidence lower bound (ELBO). The ELBO is the sum of two loss terms: a reconstruction and a regularisation loss. The former encourages the network to accurately reconstruct the input data. The latter, on the other hand, encourages the latent space to match the chosen distribution $p(\mathbf{z}|\mathbf{x})$ as accurately as possible. In the case where $p(\mathbf{z}|\mathbf{x})$ is given by \textit{independent} unit Gaussians, the regularisation term also encourages disentanglement (i.e., uncorrelated latent variables). The standard formulation of the ELBO is:
\begin{equation}
    \begin{aligned}
        {\rm ELBO} &= L_{\rm rec}(\mathbf{x}, \mathbf{x'}) + L_{\rm reg}(p(\mathbf{z}|\mathbf{x}), p(\mathbf{z})) \\
        &= L_{\rm rec}(\mathbf{x}, \mathbf{x'}) + \beta\  {\rm KL}(p(\mathbf{z}|\mathbf{x}), q(\mathbf{z}))
    \end{aligned}
\end{equation}
with $\beta = 1$ and $KL$ representing the Kullback–Leibler \citep[KL][]{kullback_information_1951} divergence between the latent distribution $p$ and the prior $q$.

\begin{figure*}
    \centering
    \includegraphics[width=\textwidth]{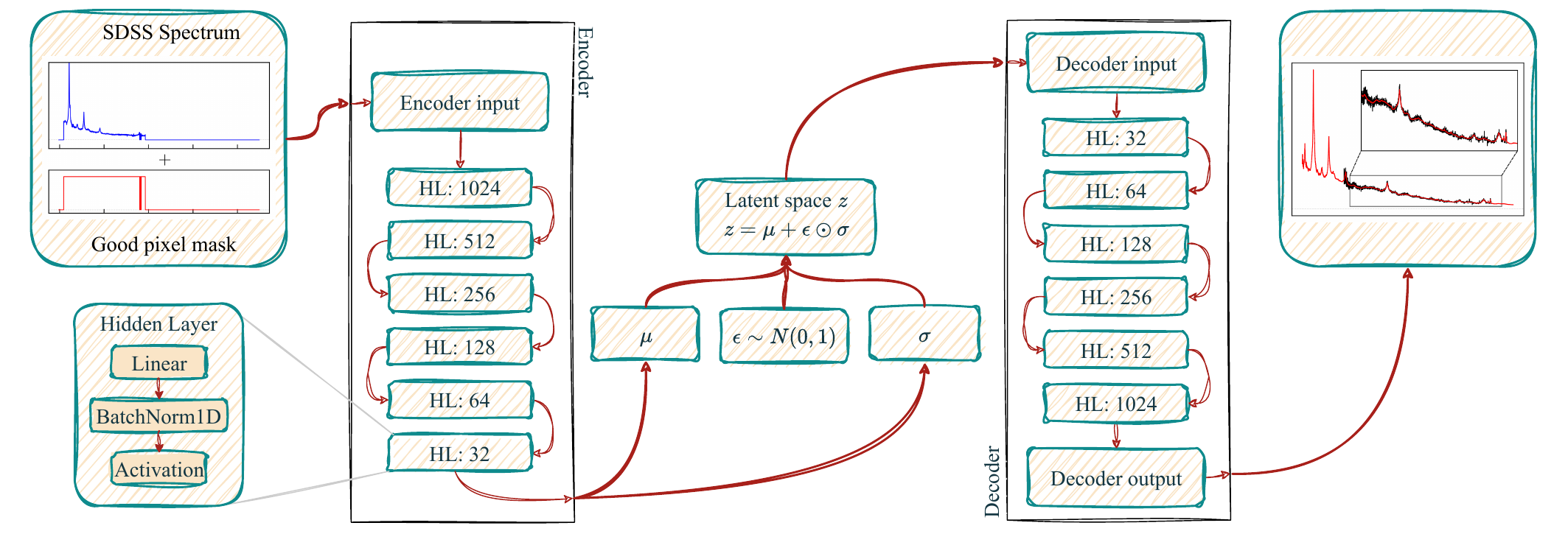}
    \caption{Schematic representation of the model architecture, input and output. The network receives as input the concatenation of an SDSS spectrum, normalised by the median spectrum, and the coverage mask (top left panel). It then encodes the input to produce a latent space representation $\mathcal{Z}$, that is decoded to produce a new spectrum (top right panel). The new spectrum (shown in red) covers a wider wavelength range compared to the corresponding input (shown in black) and is generally less noisy. The encoder and decoder are built as the reverse of each other by combining several hidden layers, denoted in the schematic by {\rm HL} followed by the corresponding output dimension. Each hidden layer is a combination of a linear layer, followed by a \texttt{BatchNorm1D} layer and the activation function (bottom left).}
    \label{fig:VAE_arch}
\end{figure*}

Several variations of this standard picture have been proposed in order to address issues with the classic VAE implementation. Examples include $\beta$-VAEs \citep[][where $\beta \neq 1$]{higgins_beta-vae_2017} and InfoVAEs \citep{zhao_infovae_2017}, which we employ in this work. Two main reasons motivated the introduction of the InfoVAE: on the one hand, the regularisation part of the loss function can be too strong with respect to the reconstruction; on the other, ELBO-based VAEs tend to overfit the data if the training dataset is not sufficiently large. In practice, both issues result in a VAE that does not learn a meaningful representation of the data either because the algorithm simply produces $q(\mathbf{z})$ regardless of the input or because it overfits the data without actually learning the underlying distribution.
An InfoVAE addresses this issue by modifying the loss and including an additional term
\begin{equation}
    \begin{aligned}
        \label{eq:infoVAE_full_loss}
        L_{\rm InfoVAE} = L_{\rm rec}(\mathbf{x}, \mathbf{x'}) + (1 - \alpha)\ {\rm KL}(p(\mathbf{z}|\mathbf{x}), q(\mathbf{z})) + \\ (\alpha + \lambda - 1)\ {\rm MMD}(p(\mathbf{z}|\mathbf{x}), q(\mathbf{z}))
    \end{aligned}    
\end{equation}
where MMD represents the Maximum Mean Discrepancy \citep[MMD, ][]{gretton_kernel_2012}, computed between each latent space dimension $\mathbf{z}$ and the prior $q(\mathbf{z})$. This new loss addresses both issues: on the one hand, the strength regularisation term can be lowered, tailored to specific applications, or removed altogether. The additional regularisation term, based on the MMD, encourages a better use of the latent space and has been shown to be significantly less prone to overfitting \citep{zhao_infovae_2017}.

\subsection{Model architecture, training strategy and hyperparameters}
\label{sec:optimisation}
A schematic representation of our InfoVAE architecture is shown in Fig. \ref{fig:VAE_arch}. We employ a symmetric architecture, in which the encoder and decoder mirror each other. The network receives as input the concatenation of the preprocessed spectra (divided by the median spectrum, as we found this to make the training more stable) and the respective coverage mask, added to explicitly inform the network about the wavelength range covered by each spectrum, and whether a given pixel should be ignored for any reason. The concatenation is passed through a series of hidden blocks to produce two vectors, $\boldsymbol{\mu}$ and $\boldsymbol{\sigma}$. Through the reparametrisation trick, these are encoded in the latent space $\mathcal{Z}$ and finally decoded to produce the reconstructed spectrum. Each hidden block is constituted by a linear, fully connected layer followed by batch normalisation and the activation function. We opt for the activation function proposed by \citet{2020ApJS..249....5A}, which we found to outperform the widely used LeakyReLu \citep[Leaky Rectifier Linear unit,][]{maas_rectifier_2013}. The network is implemented in PyTorch \citep[version 2.7][]{ansel_pytorch_2024} and trained using the Adam optimiser \citep{kingma_adam_2017}. The reconstruction loss is defined as the $\chi^2$ statistic between the reconstructed and corresponding input spectra, computed using the formal SDSS inverse variance. This has the advantage of naturally taking into account the uncertainty in the training data, which would otherwise be ignored. We follow the standard InfoVAE implementation for the regularisation term, but set $\alpha = 0$ in Eq. \ref{eq:infoVAE_full_loss}, following the recommendation of \citet{zhao_infovae_2017}. We identify an optimal $\lambda$ by hyperparameter optimisation (Table \ref{tab:params_optimised}).
\begin{figure}[ht]
    \centering
    \includegraphics[width=\columnwidth]{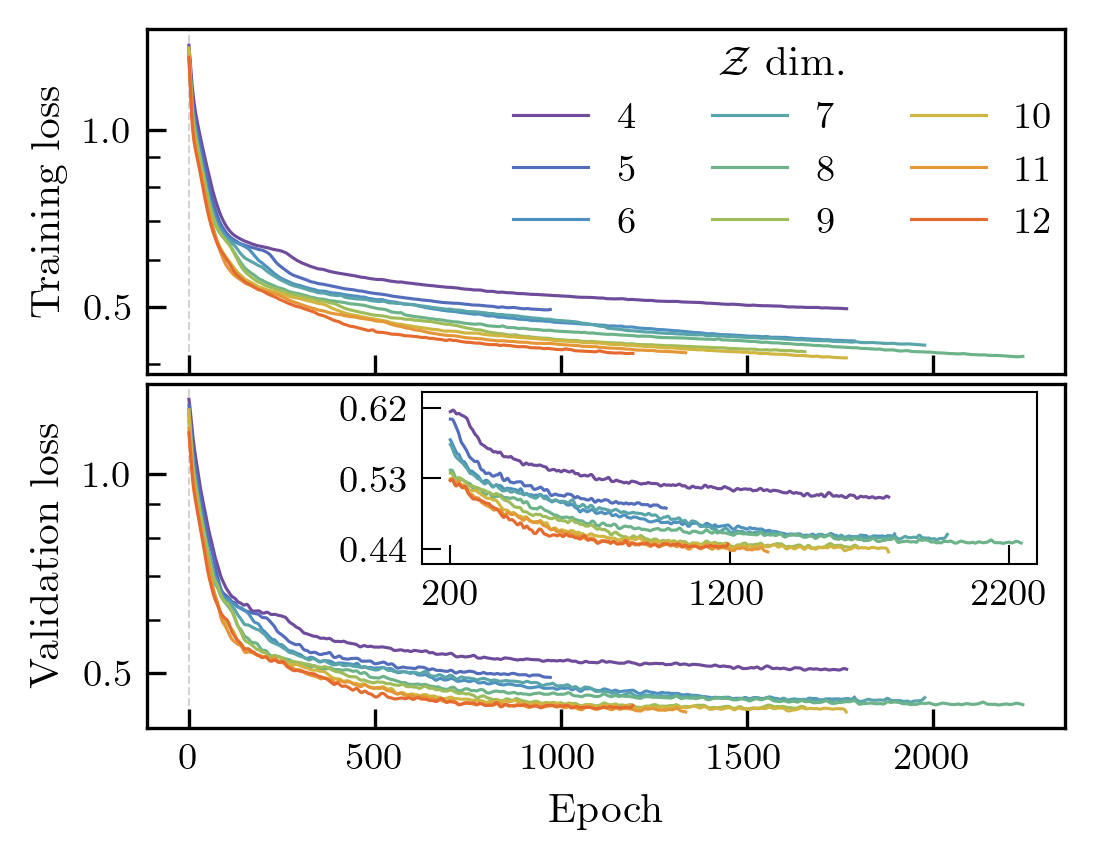}
    \caption{Training and validation losses for the GP network as a function of the number of latent dimension. It is clear from the lower panel that further increasing the number of latent dimensions beyond ten does improve in the validation loss.}
    \label{fig:training_validation_loss}
\end{figure}
\begin{figure*}[b]
    \centering
    \includegraphics[width=\textwidth]{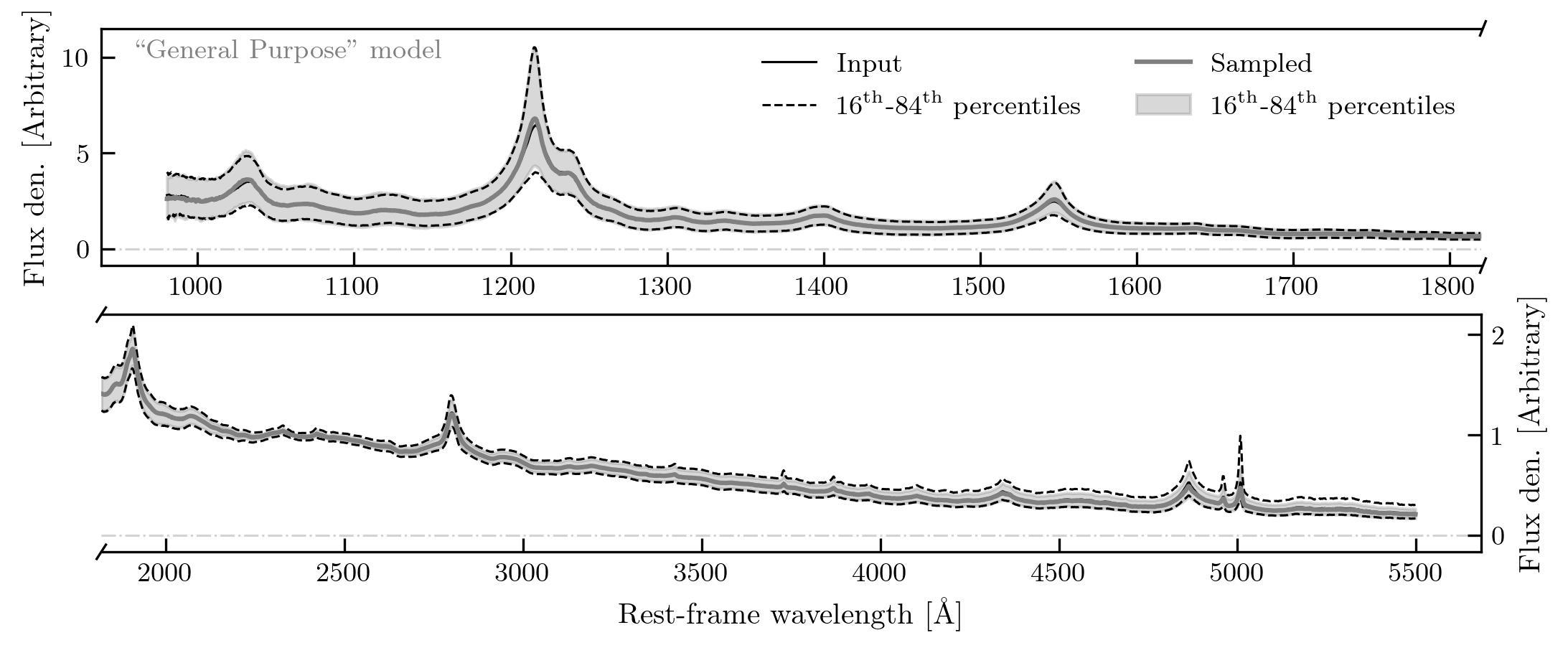}
    \caption{Sampled spectra from the GP model compared to the input spectra. In both panels, the black solid and dashed line indicate the median, 16$^{\rm th}$ and 84$^{\rm th}$ percentile of the input data. The solid, grey line represents the median spectrum of 10~000 realisations sampled from the VAE, while the shaded area encompasses the 16$^{\rm th}$ and 84$^{\rm th}$ percentile of the same sampled data. The model is able to accurately reproduce both the median and the variance of the input spectra. We note that all spectra are normalised using the flux between 2350~\AA{} and 2360~\AA{} as reference. The comparison between median sampled spectrum and median input for the FOR and FOB models is shown in Fig. \ref{fig:FOB_FOR_sampled}.}
    \label{fig:sampled_vs_input}
\end{figure*}
To limit overfitting, at run-time, we randomly mask out part of the spectra before feeding them as input to the encoder. This mask is not considered when computing the reconstruction loss. This strategy is commonly employed in denoising AEs to encourage the model to learn intrinsic and robust properties of the population. We use a batch size of 128 and train the network for 5000 epochs, but implement an early stopping strategy to interrupt the process if the validation loss does not improve for more than 200 consecutive epochs. The training and validation losses for the GP network are shown in Fig. \ref{fig:training_validation_loss}, as a function of the latent dimension.
\begin{table}[t]
    \centering
    \caption{Parameters optimised as part of the grid search. For each parameter, we list the lower, upper bound, step size and, when appropriate, whether we use a linear or logarithmic grid.}
    \begin{tabular}{c|c|c|c}
        \toprule
        Parameter & Searched interval & Type & $\Delta$ \\
        \midrule
        \# of latents       & 4 -- 12                                  & Linear & 1   \\
        $\lambda$           & $10^{-5}$ -- $10$                        & Log.   & 10  \\
        Loss                & RMSE or $\chi^2$                         & --     & --  \\
        Act. func.          & LeakyReLu or \citet{2020ApJS..249....5A} & --     & --  \\
        \bottomrule
    \end{tabular}
    \label{tab:params_optimised}
\end{table}
\begin{table}[t]
    \centering
    \caption{Parameters used to train the best model after the optimisation procedure.}
    \begin{tabular}{c|c|c|c|c}
        \toprule
        Model & \# of latents & $\lambda$ & Loss     & Act. func. \\
        \midrule
        GP    &     11        &     0.1   & $\chi^2$ & \citet{2020ApJS..249....5A} \\
        FOR   &     9         & 10$^{-5}$ & $\chi^2$ & \citet{2020ApJS..249....5A} \\
        FOB   &     9         & 10$^{-4}$ & $\chi^2$ & \citet{2020ApJS..249....5A} \\
        \bottomrule
    \end{tabular}
    \label{tab:best_params}
\end{table}
It is evident that employing a number of latent dimensions larger than ten does not improve the validation loss. We use this to limit the grid of parameters we search in the optimisation step. To optimise the hyperparameters of the network, we select a limited subset of them, listed in Table \ref{tab:params_optimised}, and perform a systematic grid search. We do not vary all possible parameters and do not change the architecture in order to keep the run-time of test runs manageable. We select the best network as the one that provides the best reconstruction. We adopt the same architecture, and optimise the hyperparameters in the same way, for all training sets.

The output of the ``best'' model for the GP dataset is shown in Fig. \ref{fig:sampled_vs_input} (and the equivalent for the FOR and FOB datasets in Fig. \ref{fig:FOB_FOR_sampled}). Here, we sample spectra from the InfoVAE and compare them with the input data. In particular, we show with the solid black line the median input spectrum and with the solid grey line the median sampled spectrum. The dashed black line encloses the 16$^{\rm th}$--84$^{\rm th}$ percentile of the input data, whereas the grey-shaded area the 16$^{\rm th}$--84$^{\rm th}$ percentile of the sampled spectra. All models show excellent agreement with the input data, both in terms of median spectrum and variance. The emission lines are faithfully reproduced, as is the quasar continuum. The variance is reduced to almost zero at $\sim2350~\angstrom$: this is expected, as it is the window in which we normalise the spectra.

\section{Latent space exploration}
\label{sec:latent_space_exploration}
After training each model, we explore the properties of the latent space to understand whether the latent dimensions reflect a particular (or a combination of) quasar physical properties. We employ different methods, both exploratory and well established, and focus our analysis on the GP model. We first explore the latent space variations through visual inspection, varying one latent space dimension at a time while keeping the others constant. We then decode each mock latent space representation and observe the effect of each latent on the reconstructed spectrum. Secondly, we apply an unsupervised dimensionality reduction algorithm \citep[Uniform Manifold Approximation and Projection for Dimension Reduction, hereafter UMAP,][]{2018arXiv180203426M} to the latent space, projecting it onto a two-dimensional embedding. We then colour code the representation and look for trends and clusters. Finally, we compute the Mutual Information \citep[MI][]{shannon_mathematical_1948} between each latent space dimension and selected physical properties of the SDSS quasars derived in \citet{2022ApJS..263...42W}. To do so, we employ \texttt{GMM-MI} \citep{Piras23}, a Gaussian mixture model estimator for MI.

\subsection{Latent space variations}
\label{sec:latent_space_variations}
\begin{figure}[ht]
    \centering
    \includegraphics[width=\columnwidth]{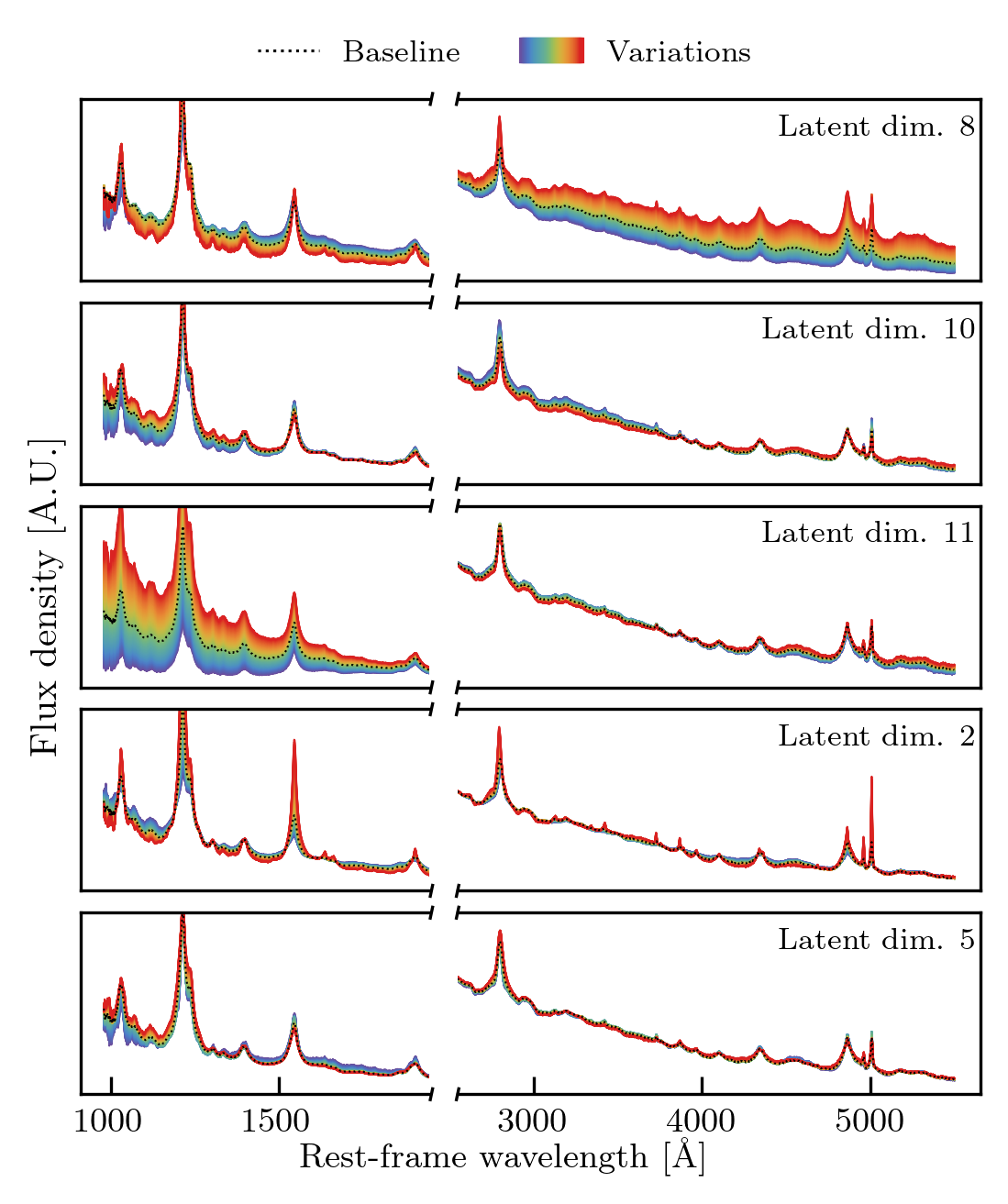}
    \caption{Decoded spectra obtained from a mock latent space where we vary a single latent dimension (indicated in the top right) while keeping the others constant. In order to better visualise the results we use a different scale for the blue and red side of the decoded spectra; the two however smoothly join. We show in this figure for the five latent space dimensions that produce the largest variation and in Fig. \ref{fig:latent_space_variations_all} all the latent dimensions.} %
    \label{fig:latent_space_variations}
\end{figure}

We initially adopt an exploratory approach to investigate whether our latent space correlates with any physical quasar property. We start by encoding the full training dataset and obtain its latent space representation. By exploiting the fact that our latent dimensions are approximately Gaussian (or, equivalently, that the mean and median of each dimension are approximately zero, see Fig. \ref{fig:latent_space_dims_corner}), we generate a ``baseline'' latent space, where each sample is represented by a vector of zeros. We expect this latent space to be close to the median quasar spectrum used to train the model (Fig. \ref{fig:medianCompositeSpec}). From this ``baseline'' latent space, we vary each latent space dimension between the respective first and 99$^{\rm th}$ percentiles while keeping the other dimensions fixed at zero. We then decode the mock latent space and plot the resulting spectra. The results, for the five latent space dimensions that produce the largest variation, are shown in Fig. \ref{fig:latent_space_variations}; the remaining are presented in Fig. \ref{fig:latent_space_variations_all}.
However, we emphasise that unlike methods such as Principal Component Analysis (PCA), the cardinality of the latent dimensions does not correlate with the amount of available information: for example, LD1 does not necessarily contain more information than the other latent dimensions. Each dimension does not capture a single spectral feature, but rather a combination of several. For example, there is a clear correlation with emission line strength (LD2, LD8, LD10 and to some extent LD5), the continuum slope (LD11) or the \FeII{} emission complex and pseudo-continuum (LD2, LD5). Emission line variations are not uniform, with some latent dimensions more evidently affecting rest frame UV or optical lines: for example, in LD10 there are significant changes in \Civ{} and \MgII{}, which are not reflected in the \Hb{} line.

\subsection{UMAP dimensionality reduction of the latent space}
\label{sec:UMAP_dim_reduction}
A more robust approach to interpreting the latent space of a VAE is to further reduce its dimensionality through dimensionality reduction algorithms, such as UMAP.
UMAP is an unsupervised and non-linear dimensionality reduction algorithm that attempts to learn the manifold structure of the data it is applied on. It produces a low-dimensional embedding that preserves the essential topological structure of that manifold \citep{2018arXiv180203426M}. Intuitively, UMAP first creates a topologically equivalent, high-dimensional representation of the data, then optimises a low-dimensional equivalent to match it, using cross-entropy as a measure of similarity. UMAP uses randomness in computing the embedding: as a consequence, the distance between clusters or the absolute values associated with each embedding point are meaningless and not deterministic. Instead, the focus should be on the resulting clusters, which reflect actual patterns in the data.

\begin{figure}[ht]
    \centering
    \includegraphics[width=\columnwidth]{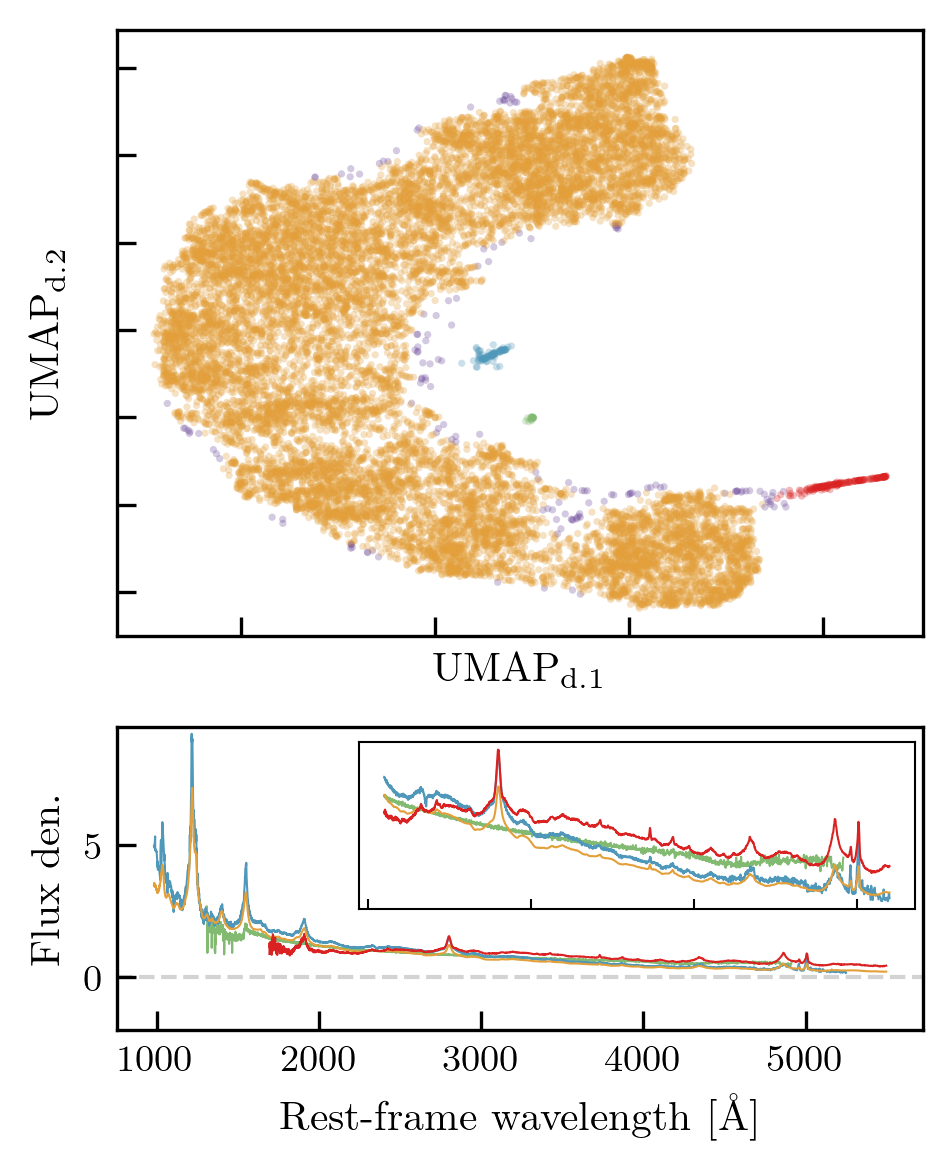}
    \caption{Two dimensional UMAP embedding of the VAE latent space, for the GP model. Top panel: the clusters identified in the embedding, highlighted using different colours (orange, green and red). Outliers (that is, points that are not associated with any cluster) are shown in purple. For visualisation purposes, the bottom panel shows the median SDSS input spectrum for objects in each cluster, with matching colour-coding.}
    \label{fig:HDBSCAN_clusters_spectra}
\end{figure}
We start from the same latent space representation obtained in the previous step and preprocess it to scale all dimensions using a \texttt{RobustScaler} from \texttt{scikit-learn}. We then fit a UMAP model to the scaled latent space representation and obtain a two-dimensional embedding of our $\mathcal{Z}$. We keep all UMAP parameters at their default values, with the exception of \texttt{n\_neighbors} (set to 15) and \texttt{mid\_dist} (set to 0.01). We determine these values through trial and error: the embedding results do not significantly depend on the choice of hyperparameters as long as \texttt{n\_neighbors} is not too large ($\gtrsim$ 50). Finally, for visualisation purposes and qualitative analysis, we apply a clustering algorithm \citep[\texttt{HDBScan},][]{McInnes2017} to automatically identify clusters in the UMAP embedding. The results are shown in Fig. \ref{fig:HDBSCAN_clusters_spectra}.
\begin{figure}[ht]
    \centering
    \includegraphics[width=\columnwidth]{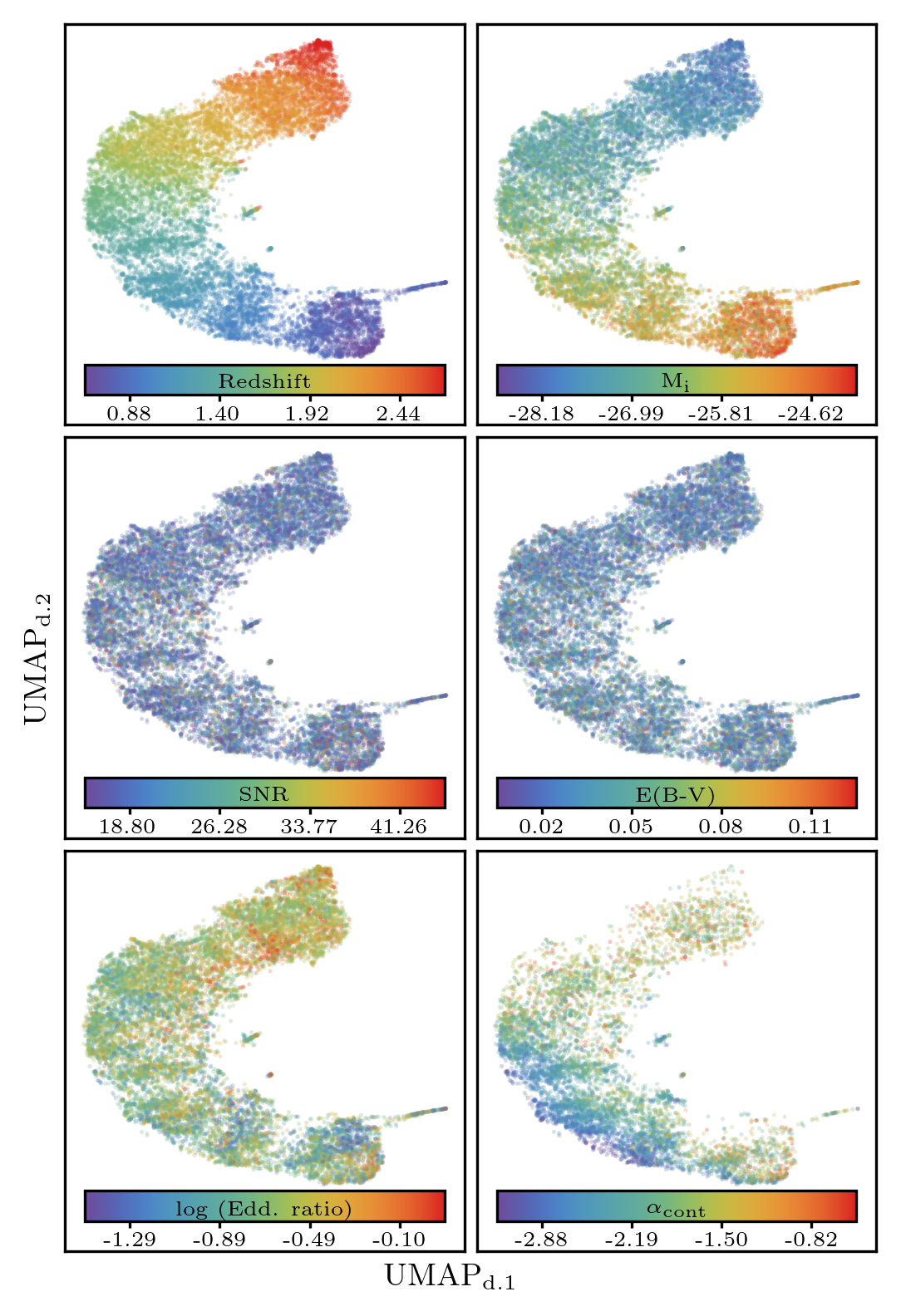}
    \caption{UMAP embedding colour-coded by redshift, absolute \textit{i}-band magnitude, S/N, galactic extinction, logarithm of the Eddington ratio and continuum slope (top to bottom, left to right).}
    \label{fig:UMAP_color_coding_subset}
\end{figure}
\begin{figure*}
    \sidecaption
    \includegraphics[width=12cm]{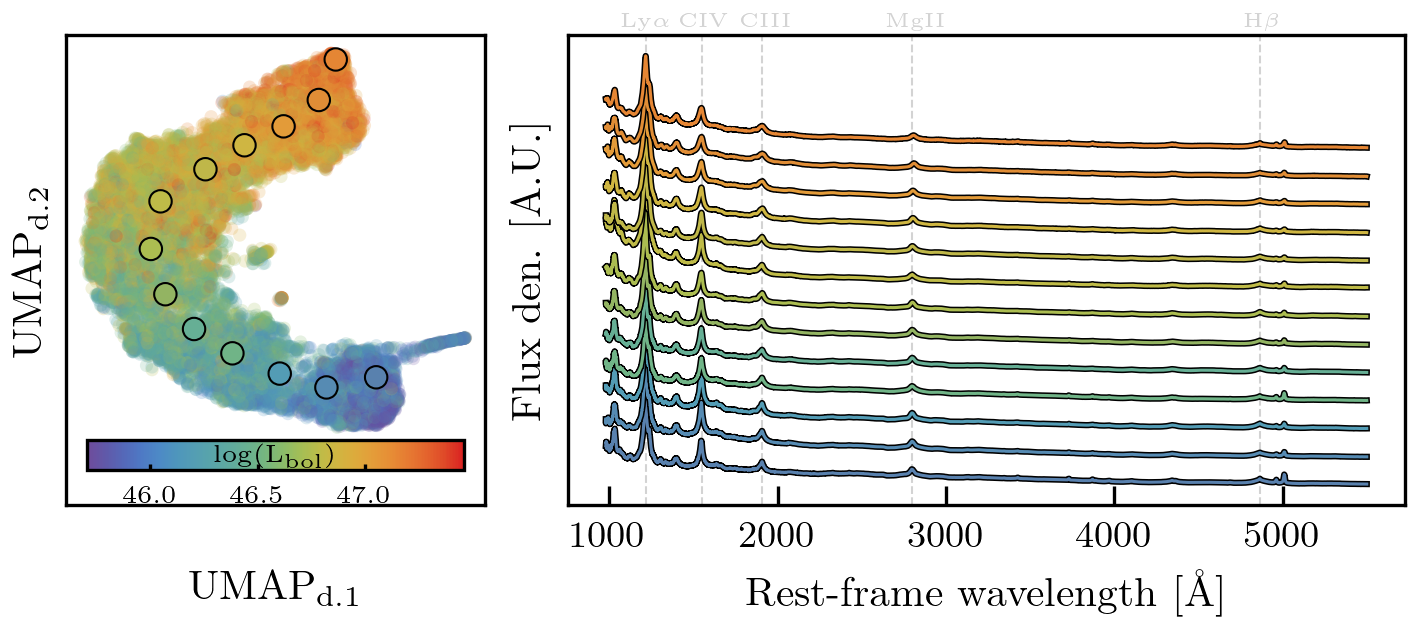}
    \caption{UMAP representation colour-coded as a function of the bolometric luminosity, and resulting spectra decoded from latent space point along the direction of evolution. Left panel: UMAP embedding; the points on which we apply the inverse UMAP transformation are highlighted as scatter points and colour-coded using the average neighbouring colour. Right panel: spectra decoded from latent space samples highlighted in the left panel and colour coded according to the respective originating scatter point.\vspace*{0.9cm}}
    \label{fig:UMAP_evolution_decoded_spectra}
\end{figure*}
The UMAP embedding features a smooth and large cluster (``main'', orange), two small clusters (blue and green), and an extended tail (red). This reflects the homogeneity of the training set, designed to be as clean as possible of reddened spectra, spectra with BALs, and artefacts. As shown by each median spectrum in the bottom panel of Fig. \ref{fig:HDBSCAN_clusters_spectra}, spectra belonging to the ``main'' cluster are the closest to typical type-1 quasars. Spectra belonging to the red ``tail'' are redder than the typical quasar and exclusively at low-$z$, whereas those in the blue cluster appear bluer than the average. Finally, spectra belonging to the green clusters lack most of the typical quasar emission lines. This could indicate that the VAE has learnt to recognise blazars, spectra misclassified as quasars, or spectra with an incorrect redshift. We visually inspect each of them (twenty in total), noting that 75\% do not show prominent emission lines, while the remaining have not been assigned the correct redshift. These objects showcase the power of the VAE as a tool for identifying outliers and errors in large catalogues and will be removed in the future from all datasets.

Furthermore, we plot the resulting UMAP embedding and colour-code each point according to selected quasar properties derived in \citet{2022ApJS..263...42W}. The results are shown in Figure \ref{fig:UMAP_color_coding_subset}.
The S/N and the galactic reddening are not correlated with the UMAP embedding: this implies that the model did not learn the noise pattern of the SDSS spectra and that the de-reddening applied during the preprocessing successfully removed the effect of galactic extinction. Instead, there is a strong gradient in redshift and absolute \textit{i}-band magnitude. This trend can be attributed either to the addition of the coverage mask as input to the model, or to selection effects inherited from the SDSS survey (Fig. \ref{fig:MI-z_dist}), or to a combination of both. We colour-code the last panel by the logarithm of the Eddington ratio, to check whether the model has learnt a physically meaningful quantity. The result hints towards a positive answer, as it is possible to identify regions of the embedding where quasars with high or low Eddington ratios are grouped together.

Finally, we investigate how the reconstruction changes as a function of the coordinate in the UMAP embedding. To do so, we employ the \texttt{inverse\_transform} method implemented in UMAP and follow the bolometric luminosity trend as illustrated in Fig. \ref{fig:UMAP_evolution_decoded_spectra}. We arbitrarily place 13 points (coloured circles, left panel) following the change in bolometric luminosity. We then obtain the corresponding points in the latent space by applying the inverse UMAP transformation, decode them into spectra, and plot them stacked on top of each other (right panel). Several interesting trends appear. It is immediately noticeable that sampling from the region with the highest bolometric luminosity produces quasars with the weakest emission lines: this indicates that the VAE has learnt the Baldwin effect \citep{baldwin_luminosity_1977}. In addition, quasars with higher bolometric luminosity produce broader lines: this is consistent with our expectations of them having larger black hole masses. Furthermore, we check whether the peak of the most prominent emission line shifts as a function of the bolometric luminosity. Surprisingly, the peak position of \MgII{} evolves redward with redshift, whereas the \Civ{} does not evolve at all. Both behaviours are unexpected: previous works have a correlation between luminosity (or redshift, as argued by) and \Civ{} blueshift that is not observed in \MgII{}. A possible explanation is a systematic issue in the SDSS redshift pipeline for spectra in which only rest-frame UV emission lines are available. In these cases, blue-shifted \Civ{} emission could lead to an underestimated redshift estimate, in turn causing a redshifted \MgII{} line.

\subsection{Mutual Information}
To quantitatively measure the correlation between latent space dimensions and quasar physical properties, we compute the Mutual Information between each latent space dimension and selected quasar properties, again obtained from \citet{2022ApJS..263...42W}. The MI is a measure of the mutual dependence between two random variables $X$ and $Y$. It captures linear and non-linear correlation between two $X$ and $Y$ and is defined in terms of the Kullback-Leibler divergence $D_{\rm KL}$:
\begin{equation}
    {\rm MI}(X; Y) := D_{\rm KL}\left(P_{\left(X, Y\right)} || P_{X} \otimes P_{Y} \right)
\end{equation}
with ($X, Y$) being a pair of random variables defined over a space $\mathcal{X}\times\mathcal{Y}$, $P_{\left(X, Y\right)}$ their joint distribution and $P_{X}$, $P_{Y}$ their marginal distributions, and $\otimes$ denoting the outer product between the two marginal distributions. MI is, by definition, non-negative and equal to zero only when $X$ and $Y$ are completely independent.
\begin{figure}[ht]
    \centering
    \includegraphics[width=\columnwidth]{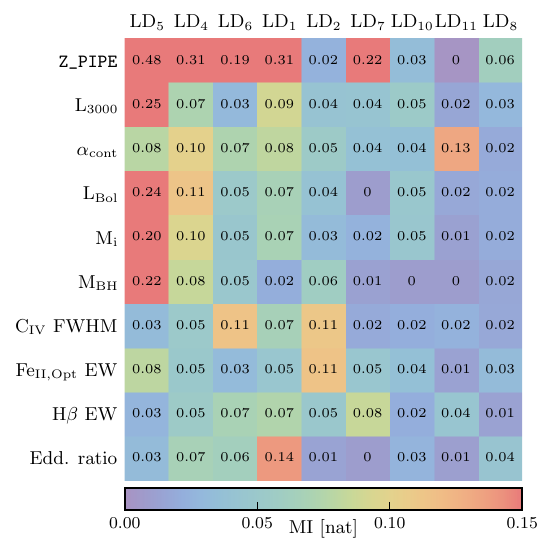}
    \caption{Mutual Information between all the latent dimensions of the GP model and the ten most correlated variables.}
    \label{fig:MI_GP}
\end{figure}
If $X$ and $Y$ are continuous random variables, then MI can be written as:
\begin{equation}
    \label{eq:MI_continuous}
    {\rm MI}_{\left(X; Y\right)} := \int_\mathcal{Y}\int_\mathcal{X} P_{\left(X, Y\right)}(x, y)\ln\left(\frac{P_{\left(X, Y\right)}(x, y)}{P_{X}(x) P_Y(y)}\right) dx\ dy
\end{equation}
with $P_{\left(X, Y\right)}$ representing the joint probability density function of $X$ and $Y$, and $P_X$ and $P_Y$ the respective marginal probability density functions. MI is expressed in ``nat'' (natural unit of information) when taking the natural logarithm of the ratio, as is in Eq. \ref{eq:MI_continuous}. Several methods have been proposed to compute the MI between two random variables, including histograms and Gaussian mixture models. In this work, we make use of the publicly available \texttt{GMM-MI} Python package \citep{Piras23}, which estimates the probability density functions using Gaussian mixture models, and has the added benefit of providing uncertainties through bootstrap resampling. The algorithm was designed and applied in the past to interpret the latent space of deep learning models \citep{lucie-smith_explaining_2024, lucie-smith_deep_2024}. A detailed discussion of \texttt{GMM-MI} is beyond the scope of this paper, and we refer the interested reader to \citet{Piras23} for a thorough description.

We present the results of mutual information analysis in Fig. \ref{fig:MI_GP}, where we show the ten most correlated properties and the corresponding mutual information values for each latent dimension (with the exception of LD3 and LD9, excluded due to their minimal correlations). As discussed in Sect. \ref{sec:latent_space_variations}, most latent dimensions correlate with several variables. LD5, in particular, strongly correlates with bolometric luminosity, BH mass, continuum luminosity, and M$_{\rm i}$. As noted in Sect. \ref{sec:UMAP_dim_reduction}, these correlations probably originate from the SDSS selection function. LD2, on the other hand, correlates with emission line properties, as do LD1 and LD6. Most dimensions also show a significant correlation with redshift and with the continuum slope, particularly LD11, consistent with our findings from Fig. \ref{fig:latent_space_variations} and Fig. \ref{fig:UMAP_color_coding_subset}.

\section{Applications}
\label{sec:VAE_application}
In this section, we showcase the capabilities of the trained VAE models to perform a variety of tasks, from generating quasar photometry to reconstructing quasar emission lines in order to compute their black hole masses. In all cases, we will use the ``best'' model trained on a specific dataset, where ``best'' is defined as in Sect. \ref{sec:optimisation}.

\subsection{Generation of synthetic quasar photometry}
\label{sec:synthetic_photometry}
The most straightforward application of the GP model (and the initial goal that motivated the development of this Info-VAE) is the generation of quasar photometry, given a redshift range $[z_{min}, z_{max}]$ and a reference absolute magnitude range [$M_{1450, min}, M_{1450,max}$]. For this application, the GP model is optimal, featuring the largest wavelength coverage, thus allowing the most flexibility in generating photometry in different filters.

We start by sampling the latent space and generate synthetic quasar spectra. They cover the rest-UV and optical wavelength range from 980~\AA{} to 5500~\AA{}. We do not bias the sampling towards any quasar spectral property besides those that originate from the training set. Because of this, the sampling naturally reflects the diversity of quasar spectral shapes captured by the SDSS quasar sample, without the need to explicitly model them. Although realistic in terms of spectral shape, the examples generated by the VAE are scaled to arbitrary units and at redshift $z = 0$: as such, they need to be preprocessed before being suitable for the generation of photometric data.

We first define a reference redshift $[z_{min}, z_{max}]$ and absolute magnitude [$M_{1450, min}, M_{1450,max}$] interval, together with the number of quasars to generate. Given these priors, we sample the $z-M_{1450}$ space and produce tuples ($z_{i},\ M_{1450, i}$). The sampling can be either uniform, according to a user-defined quasar luminosity function, or based on an empirical distribution estimated from user-provided data.
We then smoothly join the \citet[Table 1 in the paper]{lusso_first_2015} quasar template with the generated spectra. First, we define an overlap range of 980\AA{} and 1020\AA{}. Then, we rescale the \citet{lusso_first_2015} template so that it matches the quasar pseudoflux in this wavelength window. Finally, we smoothly join the template and the sampled spectrum.

We then associate to each $z_{i},\ M_{1450, i}$ pair a quasar spectrum. Each spectrum is shifted to the assigned redshift by multiplying the wavelength axis by $(1 + z_i)$, and scaled to the respective M$_{1450}$ by first computing the apparent magnitude $m_{1450, i} = M_{1450, i} + {\rm DistMod} + K_{\rm corr}$, where ${\rm DistMod}$ represents the distance modulus computed using the standard \citet{planck_collaboration_planck_2020} cosmology and $K_{\rm corr} = 2.5 \log_{10}(1 + z_i)$. We redden each spectrum using the same reddening model employed during the generation of each training dataset \citep[][see Sect. \ref{sect:GP_dataset_prep})]{2023ApJ...950...86G}. To do so, we generate galactic coordinates $(l, b)$ by uniformly sampling $l$ between 0$^{\circ}$ and 360$^{\circ}$, and $b$ between -90 and +90, ensuring $|b| > 15^{\circ}$ for consistency with the training data set.
Finally, we use SimQSO \citep{mcgreer_simqso_2021} to generate random realisations of IGM absorption spectra. We multiply these by each sampled quasar spectrum to simulate the effect of the Lyman forest and depress the flux bluewards of the \lya{} emission line. This completes the pre-processing steps.
\begin{figure}[ht]
    \centering
    \includegraphics[width=\columnwidth]{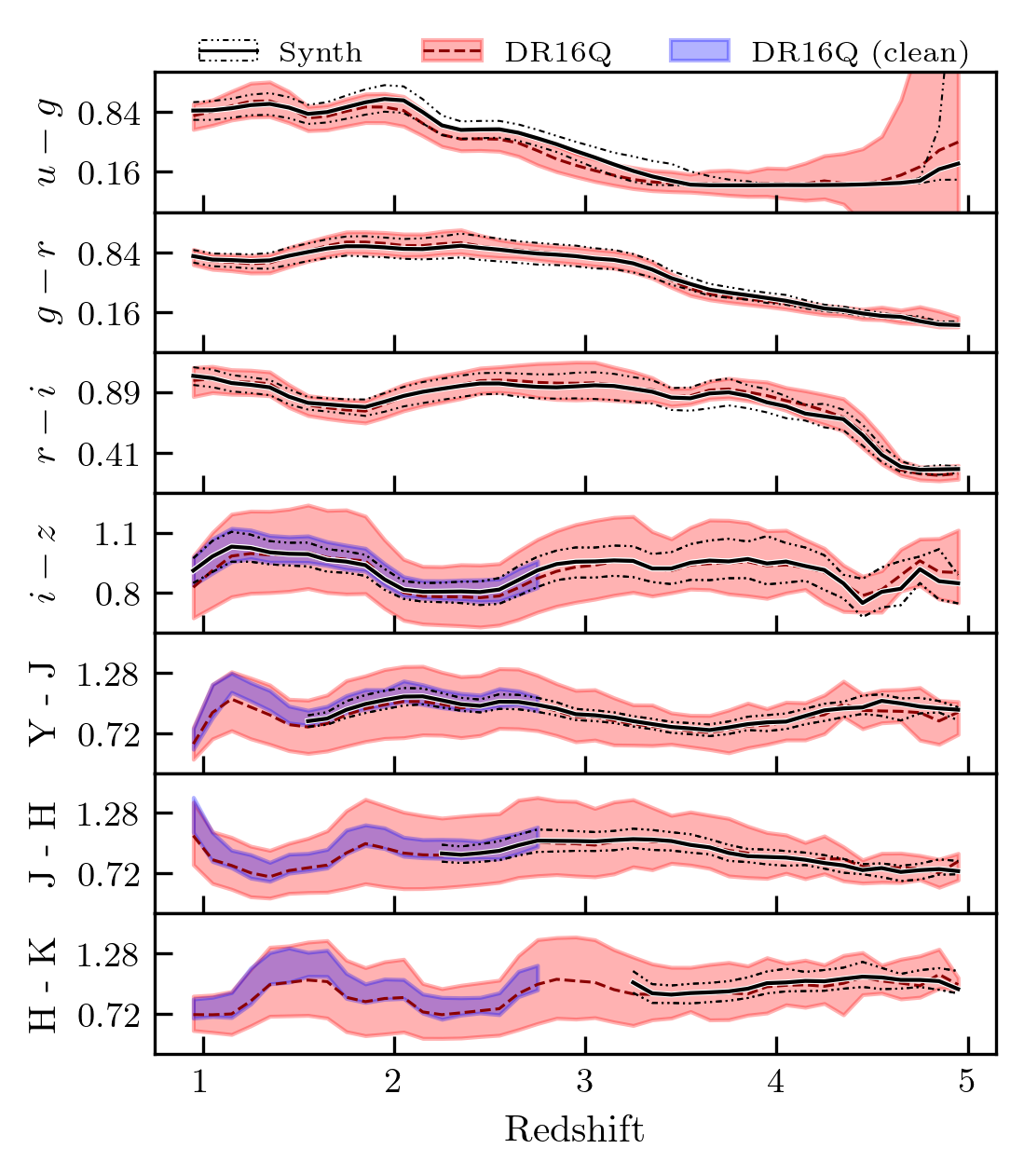}
    \caption{SDSS and UKIDSS colours as a function of redshift for the SDSS quasar and our synthetic photometry. The model faithfully reproduces the SDSS median colour at all redshifts, with the exception of the $u$ - $g$ colour, where the absorption from the IGM and the interpolation over the Lyman forest significantly affects the $u$ band. The same happens for for UKIDSS colours, albeit with a narrower spread.}
    \label{fig:SDSS_photometry_sanity_check}
\end{figure}

To estimate photometry from the spectra, we use \texttt{SpecLite} \citep{kirkby_desihubspeclite_2024}. \texttt{SpecLite} convolves each spectrum with the appropriate filter response curve to obtain AB magnitudes. These AB magnitudes are uncertainty-free and do not take into account the photometric depth of the survey. To account for this, we perturb the photometry under the assumption that, for each photometric band $b$ and apparent magnitude bin $\Delta m$, the original survey error distribution is approximately Gaussian. We verify that this approximation is reasonable and estimate the error function $\sigma(\Delta m)$ (that is, the typical uncertainty as a function of apparent magnitude). Then, for each apparent magnitude $m$, we assign an uncertainty $\sigma$ using the error function and sample a new perturbed magnitude $m_{\sigma}$ from a Gaussian N($m$, $\sigma$). These $m_{\sigma}$s, together with the associated uncertainties, represent the final product of the algorithm.

As a first sanity check, we compare our synthetic photometry against the SDSS DR16Q quasar photometry. We compute the error function as outlined in the previous section, by selecting SDSS point sources. We generate the redshift-absolute magnitude grid by sampling the corresponding distributions of the SDSS DR16Q catalogue\footnote{The SDSS DR16Q provides the absolute \textit{i}-band magnitude, which we use in place of the $M_{1450}$}.
\begin{figure}[ht]
    \centering
    \includegraphics[width=\columnwidth]{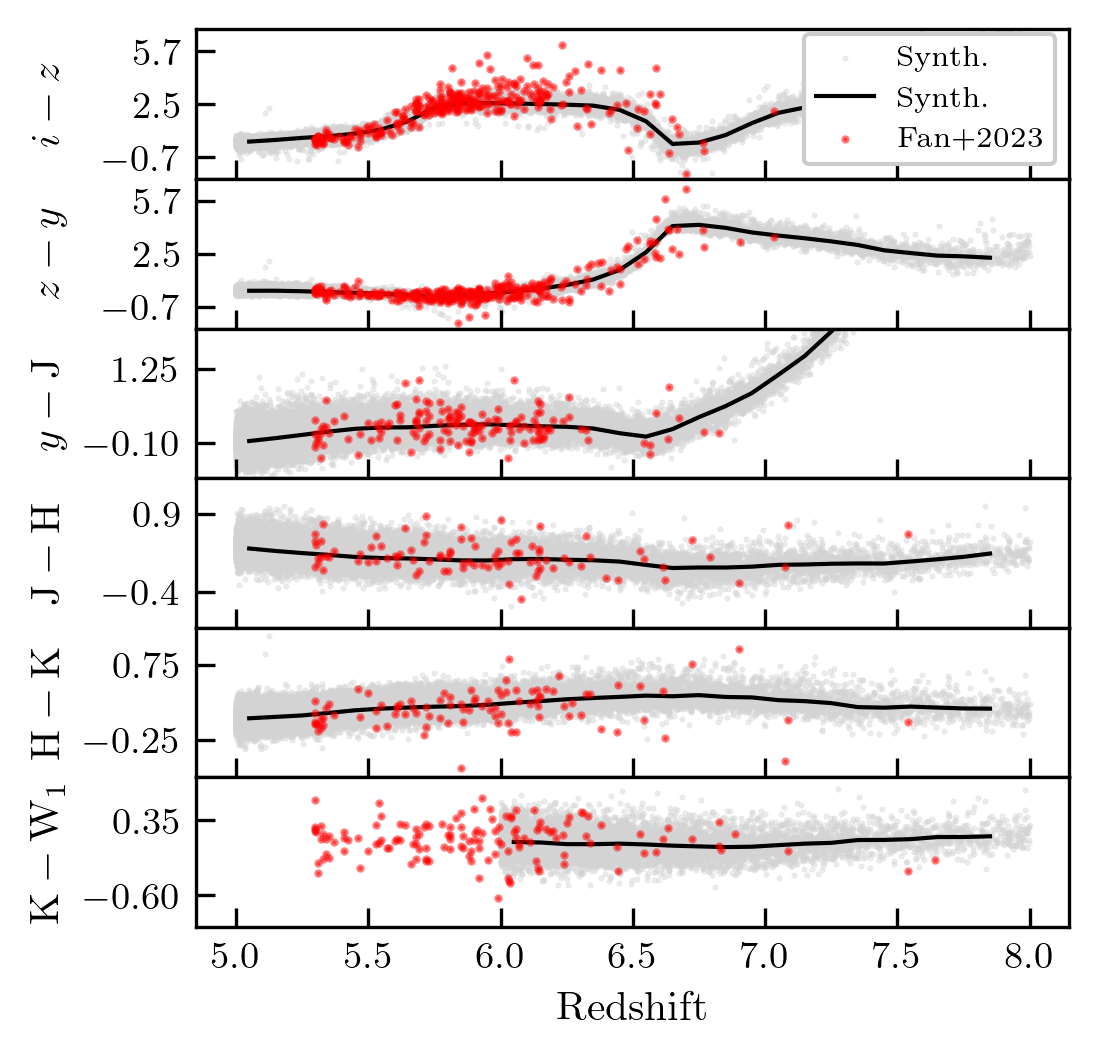}
    \caption{Same as Fig. \ref{fig:SDSS_photometry_sanity_check}, but for the high-$z$ sample from \citep{fan_quasars_2023}. In this case, we represent the photometry from the real quasars with scatter points instead of presenting the 16$^{\rm th}$-84$^{\rm th}$ percentile range due to the low number of high redshift quasars.}
    \label{fig:Synth_vs_Fan+23}
\end{figure}
The results are shown in Fig. \ref{fig:SDSS_photometry_sanity_check}, where, as a function of redshift, we show the median flux ratio from the entire SDSS DR16Q with the dashed dark red line and with the solid black line the median synthetic photometry. The red (blue) shaded area represents the 16$^{th}$ and 84$^{th}$ percentile range for the entire SDSS DR16Q (for the quasar part of the training dataset). We show the same quantities for the synthetic photometry using the dashed, black lines. In addition to the $ugriz$ SDSS photometry, we also include the UKIDDS Y, J, H, and K data. Our generated spectra do not fully cover the H and K bands at redshift $z \lesssim 3$, and as such we do not include the photometry of Fig. \ref{fig:SDSS_photometry_sanity_check}. The median SDSS, UKIDSS and synthetic photometry agree well in most cases, with some minor differences in the $u - g$ colour. This could be attributed to different factors: on the one hand, our reconstruction of the unabsorbed continuum blueward of the \lya{} forest could be imperfect; on the other hand, the IGM model we employ \citep{mcgreer_simqso_2021} is not fully representative of the IGM at these redshifts. Moreover, in the case of the UKIDSS bands, the spread in quasar colours does not match the SDSS data. This is likely a consequence of the censored training dataset that we are using and it is evident from the blue shaded area, which shows a consistently narrower spread in the training quasar colours.

In addition, to confirm that the model provides accurate photometry also in the highest redshift regime, we compare the synthetic photometry with that of the $z > 5.3$ quasar catalogue provided by \citep{fan_quasars_2023}. We crossmatch the catalogue against PanSTARRS \citep{chambers_pan-starrs1_2016}, the UKIRT Infrared Deep Sky Survey \citep[UKIDSS][]{lawrence_ukirt_2007}, the VISTA Kilo-Degree Infrared Galaxy survey \citep[VIKING][]{edge_vista_2013} and the Vista Hemisphere Survey \citep[VHS DR5][]{mcmahon_first_2013} using a 0.5 arcsecond radius. The results are shown in Fig. \ref{fig:Synth_vs_Fan+23}: as in the lower redshift regime, the model faithfully reproduces the quasar colour.

\subsection{Reconstruction of spectra with BAL features}
A second application for the model is the reconstruction of BAL absorption features in quasar spectra. We expect the model to be capable of interpolating over the absorption features and producing a faithful reconstruction of the underlying continuum. We employ the FOB model to carry out this test. We proceed as follows: starting from the 12$^{\rm th}$ data release of the SDSS quasar catalogue \citet{paris_sloan_2017}, we download the BAL quasar subset\footnote{retrieved from \url{https://data.sdss.org/sas/dr12/boss/qso/DR12Q/DR12Q_BAL.fits}}. Then, we select the quasars that satisfy the wavelength coverage conditions used to generate the FOB dataset. This is not necessary, as the model is capable of extending the spectra to bluer or redder wavelengths, but it provides a well-defined dataset of nine objects. In addition, it represents a ``best-case'' scenario, where the model has access to spectra covering the full wavelength range. We visually inspect the spectra, manually mask the absorption systems, and feed the masked spectra to the model for reconstruction.
\begin{figure}[ht]
    \centering
    \includegraphics[width=\columnwidth]{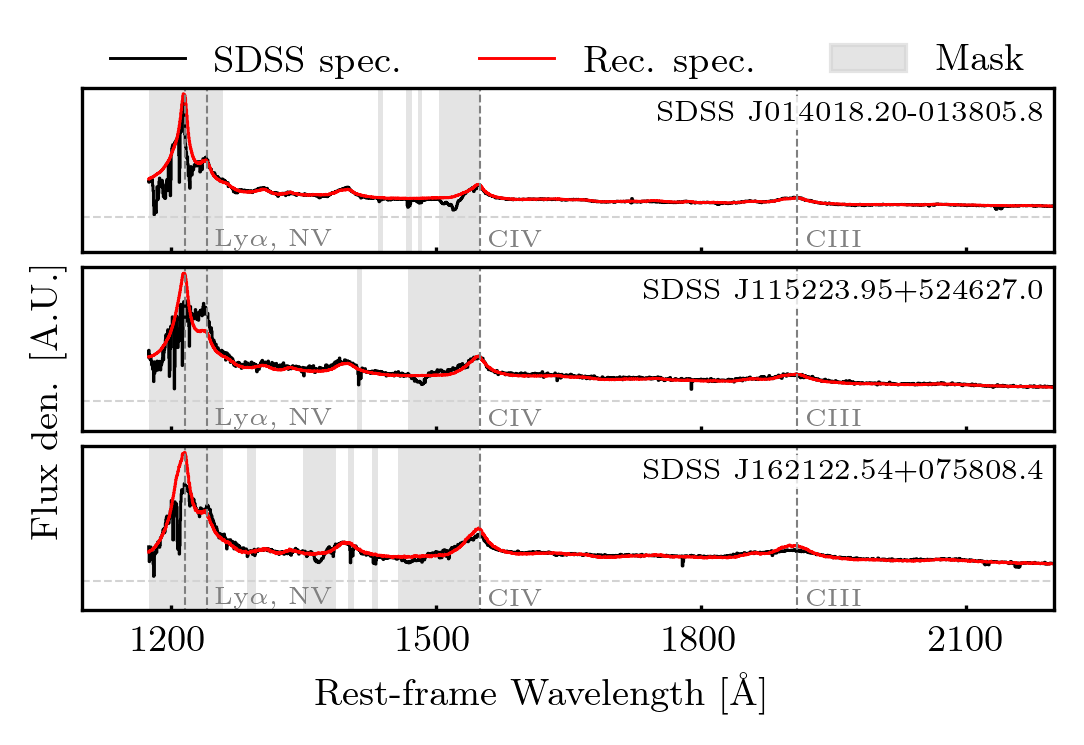}
    \caption{Example of a spectrum with an acceptable reconstruction (top panel), one where the model underestimates the \lya{} and \Nv{} complex (middle panel), and one where instead rest-frame UV and optical emission lines are not well reproduced (bottom panel). We show in black the input spectrum, in red the reconstruction and with the shaded, grey areas the masked out regions. We show the remaining spectra in Fig. \ref{fig:BAL_rec_rest}.}
    \label{fig:BAL_reconstruction_example}
\end{figure}
Qualitatively, the model reconstructs the unabsorbed continuum to a good degree of accuracy, interpolating over the absorption features and returning an unabsorbed continuum that closely matches the input spectra in most of the cases. However, the model struggles to reconstruct the emission lines, in particular the \lya{}, which appears to be often underestimated in the reconstruction, compared to the input spectra. Moreover, most of the spectra available for the reconstruction appear to feature blue-shifted components and asymmetric emission lines that the model struggles to reproduce faithfully (see, for example, the middle panel in Fig. \ref{fig:BAL_reconstruction_example}). This is hardly surprising, as it is trained on a ``clean'' dataset, devoid of spectra with similar features. Moreover, in some cases, the model underestimates the unabsorbed continuum (see again, for example, the middle panel in Fig. \ref{fig:BAL_reconstruction_example}). The reason for this is currently unclear, but a detailed exploration of this is beyond the scope of this paper.

\subsection{Black hole mass from reconstructed emission lines}
In addition to BAL quasar reconstruction, we further explore the imputation capabilities of the VAE (and, in particular, of the FOR model) and use it to reconstruct the \MgII{} and \Hb{} emission lines of selected SDSS quasars. To validate the reconstruction, we then fit the reconstructed spectra and compute the black hole mass using well established single epoch virial estimators. We finally compare the estimates with each other and with the same estimate obtained by fitting the original SDSS spectra following the approach presented in \citet{2022ApJS..263...42W}. 

In order to ensure that the test is as unbiased as possible, we only consider legacy SDSS quasars, not included in the training set. These represent the most similar but independent dataset to the spectra we used to train the algorithm. We prepare a dataset containing these spectra using the same approach outlined in Sect. \ref{sect:GP_dataset_prep}. In addition, we test different scenarios: before feeding the spectra to the VAE to reconstruct them, we either do not mask any emission line, mask only the \MgII{} or the \Hb{} emission line, or both. This serves as an additional test to check whether the model utilises information from either emission lines to compensate for the lack of the other, or if continuum information is sufficient.
\begin{figure}[ht]
    \centering
    \includegraphics[width=\columnwidth]{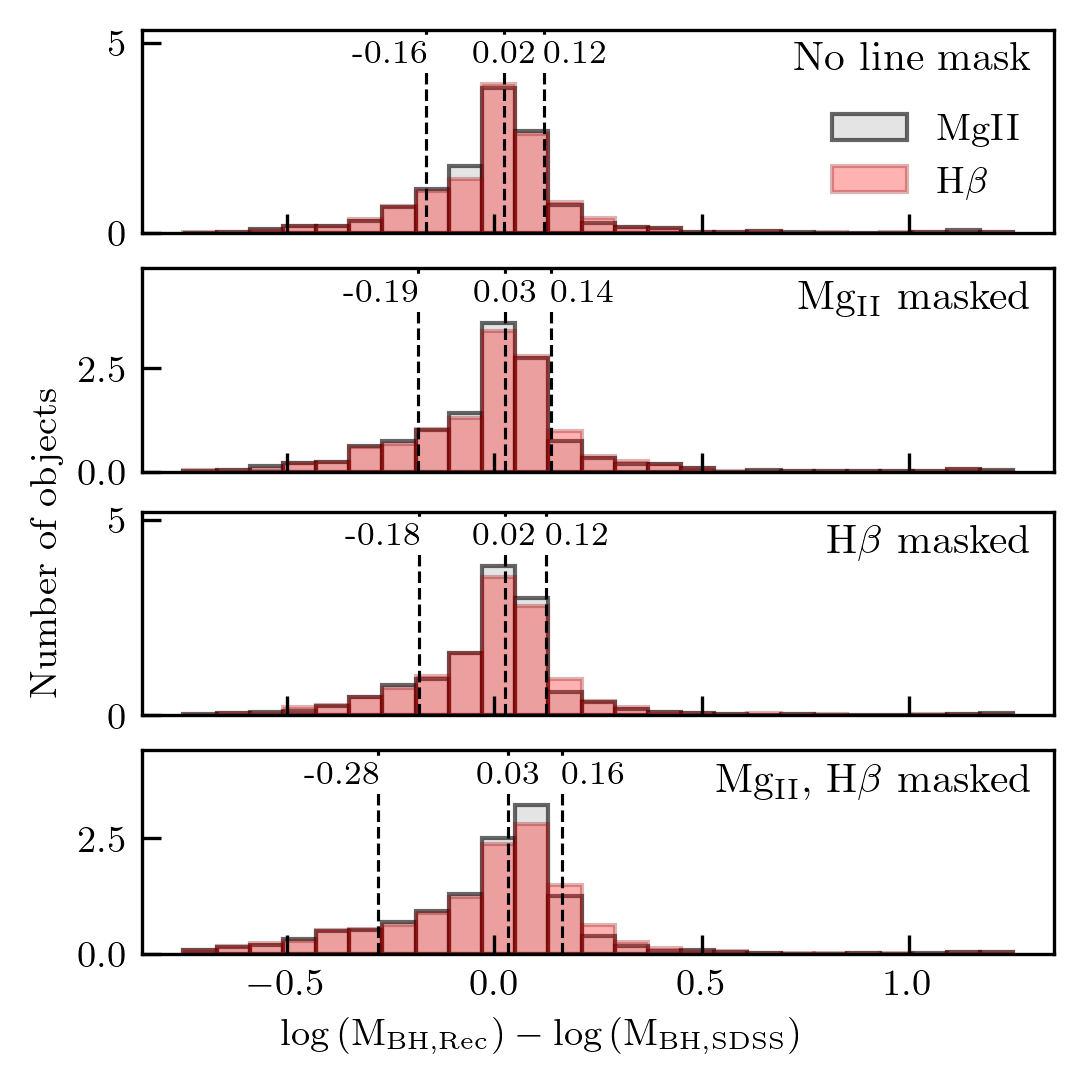}
    \caption{Logarithmic difference between the BH mass estimated from the reconstructed spectra and the original SDSS data. In all cases, the BH mass estimates are consistent and do not appear to depend on the emission line used.}
    \label{fig:BH_mass_reconstruction_comparison_relative}
\end{figure}
To be consistent with the results presented in \citet{2022ApJS..263...42W} and allow a direct comparison, we follow the same procedure and use the same input files described in their work. We present here a brief summary of the most significant steps and refer the interested reader to the original paper. All spectra were automatically modelled using \texttt{PyQSOFit} \citep{guo_pyqsofit_2018, shen_sloan_2019}. The model includes a continuum (modelled as a power law with the addition of a third-order polynomial), optical and UV \FeII{} emission using empirical templates \citep{boroson_emission-line_1992, vestergaard_empirical_2001, tsuzuki_fe_2006, salviander_black_2007} and emission lines, modelled as a combination of Gaussian profiles. Finally, for the sake of consistency and only in the case of the reconstructed spectra, we limit the fitting range to the regions originally covered by the SDSS spectra, ignoring everything else. 
The results are shown in Fig. \ref{fig:BH_mass_reconstruction_comparison_relative}, where we plot the logarithmic difference between the BH masses computed from the reconstructed and SDSS spectra. On average, the estimates of the BH masses are broadly consistent: in all cases, the median difference is close to zero. Consistently with our expectations, the best results are obtained when there is no line masking, the worst when both lines are masked out, and intermediate when only one emission line is present. This hints towards the fact that the model uses information from one emission line to reconstruct the other. It is also interesting that, especially in the case where both lines are masked, the distribution becomes more asymmetric. The larger, negative tail indicates that the BH masses computed from the reconstructed spectra tend to be underestimated compared to those derived using the SDSS spectra. This can be understood if, for example, the model struggles to reproduce the broad components of the emission lines.

\subsection{Reconstruction of the Lyman-\texorpdfstring{$\alpha$}{alpha} forest and Lyman-\texorpdfstring{$\alpha$}{alpha} emission line}
Finally, we test how well the model reconstructs the \lya{} forest and the blue side of the \lya{} emission line. We stress that, contrary to the other methods we compare against in this section, \texttt{QUEST} was not optimised for this task. However, as shown in the following, the model already performs competitively. We choose the GP model because it fully covers the required rest-frame wavelength range. 

The key idea is to reconstruct the unabsorbed quasar continuum blueward of the \lya{} emission line (1026~\AA{} -- 1210~\AA) using the unabsorbed quasar continuum redward of it (1260~\AA{} -- 2000~\AA). The reconstruction should be accurate and unbiased: both requirements are crucially important to model the unabsorbed continuum. 
This enables several scientific cases: it allows one to chronicle the end of reionisation \citep[see, e.g.,][]{bosman_hydrogen_2022} and its global timeline \citep{hennawi_precisely_2024, durovcikova_chronicling_2024}, to measure the temperature of the IGM \citep{etezad-razavi_new_2025}, or to determine the size of quasar proximity zones and to constrain quasar lifetimes \citep{Onorato2025:2505.09676v1, rojas-ruiz_first_2025}.

\begin{figure}[ht]
    \centering
    \includegraphics[width=\columnwidth]{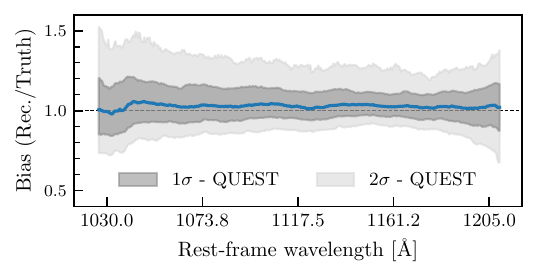}
    \caption{Bias as a function of the rest-frame wavelength in reconstructing the unabsorbed quasar continuum.}
    \label{fig:rec_quasar_continuum}
\end{figure}

In order to estimate the performance of a method, one has to choose metrics and a test set. In this work, we follow \citet{bosman_comparison_2021} and use the same dataset used to train the model, where the ``true'' unabsorbed continuum is estimated using \citet{2008A&A...491..465D} (Sect. \ref{sect:GP_dataset_prep}). We then compare the reconstruction provided by \texttt{QUEST} with the SDSS spectra. The fractional difference with respect to the truth (bias) and $16^{\rm th} - 84^{\rm th}$ percentiles range (scatter) are used as a comparison metric with other methods. We start by selecting all the quasars in the GP datasets that cover 1026~\AA{} -- 2000~\AA. Effectively, this is equivalent to restricting the comparison to the 781 quasars with \texttt{Z\_PIPE} $ \gtrsim 2.55$. For each of them, we mask out the region outside 1260~\AA{} -- 2000~\AA, feed the spectra to the VAE, reconstruct them, and compare the reconstructions with the unabsorbed continuum. The results are shown in Fig. \ref{fig:rec_quasar_continuum}, where we plot the median bias with the solid blue line and the $16^{\rm th} - 84^{\rm th}$ ($2.5^{\rm th} - 97.5^{\rm th}$) percentiles the grey (light grey) shaded regions. The reconstruction provided by \texttt{QUEST} overestimates the true continuum (with the overestimation being between 2\% and 5\%, and a median of 2.8\%). The $1\sigma$ scatter is around 10\% (+0.109/-0.092), whereas the $2\sigma$ scatter is much larger (0.301/-0.194) and strongly asymmetric (that is, the model tends to overestimate the unabsorbed continuum). Compared to the results presented in \citet{bosman_comparison_2021}, \texttt{QUEST} performs similarly to \textit{Neighbours}, outperforming \textit{Power-Law} and \textit{PCA-Pâris-10} but being outperformed by \textit{PCANN-QSANNdRA} and \textit{PCA-Davies-nominal}.

\begin{figure*}
    \centering
    \includegraphics[width=\textwidth]{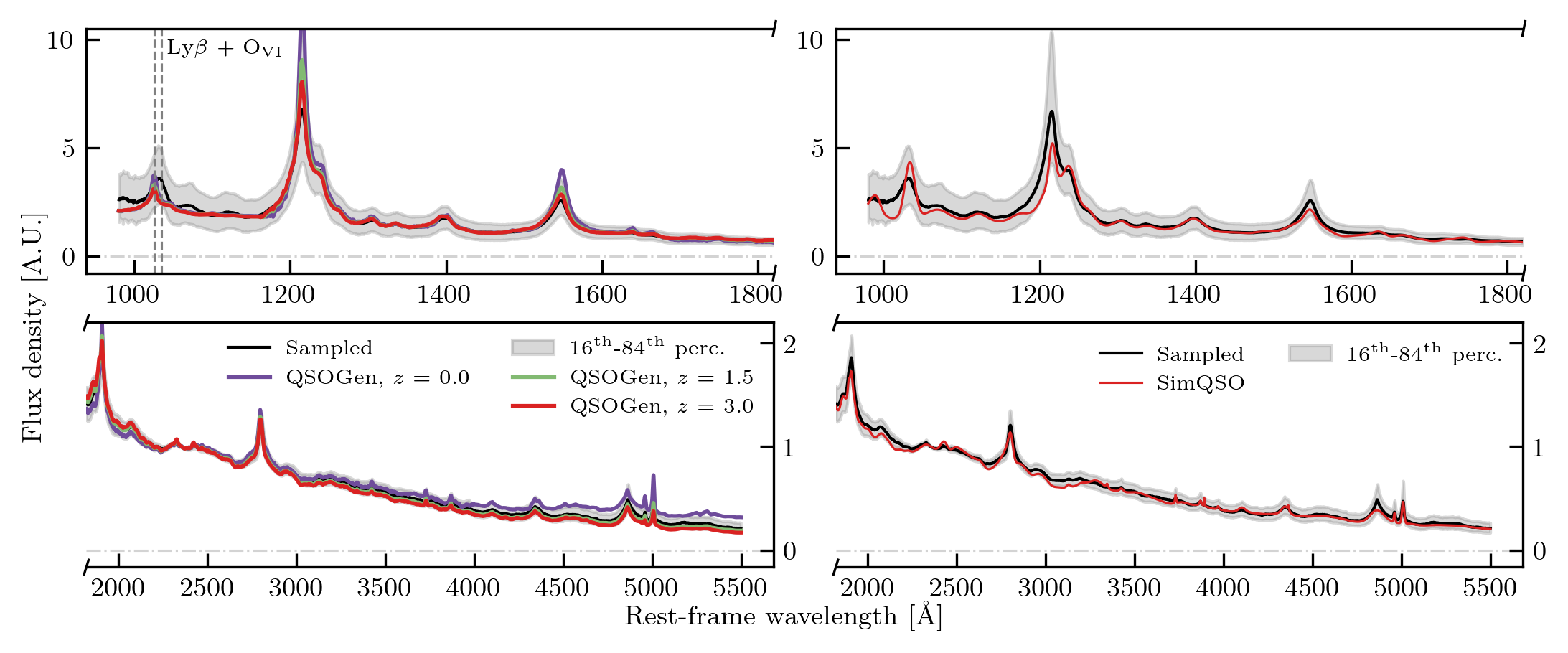}
    \caption{Median spectrum, computed from 50~000 realisations sampled from the model presented in this work, and synthetic spectra from \citet{temple_modelling_2021} (left panels) and \citet{mcgreer_simqso_2021} (right panels). In all plots the black solid line represents the median spectrum from this work with the corresponding 16$^{\rm th}$--84$^{\rm th}$ percentile range. In the two left panels, the coloured lines represent realisations of the default \texttt{qsogen} model at different redshifts, whereas in the right panels we show the median \texttt{SimQSO} spectrum with tweaked emission line strength \citep[from][]{schindler_pan-starrs1_2023} in red.}
    \label{fig:sampled_vs_temple+2021_vs_simqso}
\end{figure*}

\section{Discussion}
\label{sec:discussion}
Machine learning models are becoming increasingly common in extragalactic astronomy and in the study and search for quasars and AGNs. Among others, machine learning models have been deployed to select quasars from large photometric catalogues \citep[e.g.,][]{byrne_quasar_2024, fu_catsouth_2025}, classify optical spectra and estimate their redshift \citep[e.g.,][]{busca_quasarnet_2018, moradi_fnet_2024}, identify outliers and peculiar sources \citep{tiwari_spectroscopic_2025}, reconstruct the unabsorbed continuum leftward of the \lya{} emission line \citep[][]{2025A&A...698A.292P, hahn_reconstructing_2025}, or directly estimate the quasar's black hole mass \citep{he_predicting_2022, eilers_generative_2022}. Although many of the models mentioned above are tailored to specific tasks, the one presented here can address multiple problems effectively. Nevertheless, it is instructive to compare the spectra it generates with those produced by previous works and to point out the limitations that we are aware of: these will be addressed in future works.

\subsection{Comparison with available quasar models}
We consider the models published in \citet[\texttt{qsogen}]{temple_modelling_2021} and \citet[SimQSO]{mcgreer_simqso_2021} and qualitatively compare their median sampled spectra with our own. We show in the left two panels of Fig. \ref{fig:sampled_vs_temple+2021_vs_simqso} our median sampled spectrum in black, the 16$^{\rm th}$--84$^{\rm th}$ percentile range with a grey shaded band, and three different realisations of quasar spectra from \texttt{qsogen}, at three different redshifts: $z = 0, 1.5, 3.0$. The choice of redshifts for the QSOGen spectra is somewhat arbitrary but includes a regime where the contribution of the host galaxy is expected to be significant ($z = 0$, purple), one that is comparable to the mean redshift of our training data ($z = 1.5$, green) and one that is slightly above the maximum redshift encompassed by our training set ($z = 3.0$, red). We do not change any parameter from the default, with the exception of turning off the absorption of the IGM. This also enables us to compare our reconstruction of the Lyman forest with a completely independent approach. In general, there is a very good agreement between the \texttt{qsogen} models and our own median spectra, especially at $z = 1.5, 3.0$. At these redshifts, the most striking difference is a slightly steeper continuum slope, especially evident at longer wavelengths, and somewhat broader rest-frame UV emission lines. The latter is likely a consequence of the averaging over several thousands of realisations to produce the median spectrum, combined with the averaging across the training set performed by the model itself. In addition, it is interesting to note that bluewards of the \lya{} emission line the spectra generated using \texttt{qsogen} are relatively flat, whereas the median spectrum of our model displays several features. We regard them as real, indicative of the variety of emission features that contribute to the unabsorbed quasar continuum \citep[see for example][ for a list of the most relevant emission lines in this wavelength range]{bosman_comparison_2021}. The difference can be attributed to the approach adopted in \citet{temple_modelling_2021}, where the continuum between 970~\AA{} and 1050~\AA{} is simply extrapolated from its value at 1050\AA{}. Additionally, the \citet{temple_modelling_2021} spectrum features a much narrower \lyb{}+\OVI{} complex, with the \OVI{} emission line almost absent; in contrast, our median spectrum appears to capture this feature better, showing a broader line that peaks at an intermediate wavelength between the two emission lines. This correctly reflects theoretical expectations, where the \lyb{} emission line preferentially decays to \lya{} by emitting an H$\alpha$ photon instead of directly decaying to the ground state.

On the other hand, the \texttt{qsogen} model generated at $z = 0$ exhibits significant differences with respect to the median sampled spectrum. The emission lines are stronger, and the spectrum appears redder: both effects can be explained by considering the selection effect in the SDSS catalogue, on which the \citet{temple_modelling_2021} calibrates their model. Indeed, low-$z$ quasars have, on average, lower intrinsic luminosity than their higher redshift counterparts. Due to the Baldwin effect, this leads to stronger emission lines. In addition, the contribution of the host galaxy becomes more significant, enhancing the flux at longer wavelengths and producing redder spectra.

Comparing against \texttt{SimQSO} (Fig. \ref{fig:sampled_vs_temple+2021_vs_simqso}, right panels) is not straightforward, as \texttt{SimQSO} allow significant customisation to all spectral components. For the purpose of this comparison, we employ the same parameters as previously used in \citet{schindler_pan-starrs1_2023}. Overall, we find very good agreement between the two models, with the most notable exception being the \lya{} emission line, which appears to be stronger in our sampled composite. However, it is worth emphasising that the model presented in this paper is completely data-driven and did not require any tweaking: through training, the Info-VAE learnt to appropriately reproduce quasar spectra features without the need to introduce ad hoc, tunable parameters.

\subsection{Limitation of the model}
Despite its flexibility and capabilities, the model has some limitations that we aim to address in the future.
\begin{itemize}
    \item Limited training set: the training set we used to train the model is, by design, limited to typical SDSS type 1 quasars. Since the publication of the SDSS DR16Q catalogue, major quasar catalogues have been released, including DESI DR1 \citep{desi_collaboration_data_2025} and the nineteenth SDSS data release itself. DESI spectra could prove especially useful in improving the training set, as they would allow the inclusion of fainter targets and thus the sampling of a larger parameter space.
    Additionally, one could produce dedicated training sets to generate large samples of quasar spectra of under-represented populations.
    \item Wavelength coverage: a second, significant, limitation of the model is the limited wavelength coverage of the spectra we train the VAE on. This implies that spectra in the highest redshift bins will contribute mainly to the bluest portion of the wavelength grid, whereas the spectra at low redshift will contribute to the reddest wavelengths. Because of this, we are sceptical that the model fully captures the physical correlations between the rest-frame UV and optical properties. In this context, including NIR from \textit{Euclid} \citep[covering the wavelength range 1.21--1.89 $\mu$m, albeit with low spectral resolution of R $\sim 450$, ][]{euclid_collaboration_euclid_2023} could lead to significant improvements.
    \item Model architecture and input format: recent advances in machine learning could be incorporated in the model architecture. Several works have attempted to introduce convolutional and attention layers in AEs and VAEs, obtaining good performance and interpretable results \citep[see, for example][]{melchior_autoencoding_2023}: testing the effect of these layers in our model could be helpful in unlocking additional performance. Moreover, by design, our model incorporates a coverage mask, concatenated to each spectrum, as input. Although this is needed to inform the model about the wavelength coverage of each spectrum, it might have unwanted side effects, such as introducing or reinforcing redshift trends. An approach like that presented in \citet{hahn_reconstructing_2025} could mitigate this problem.
    \item Conditional VAE: to aid with the search of a particular type of quasar, or to understand whether quasars with particular spectral properties are systematically missed by a survey or a selection algorithm, one could condition the VAE on a given quasar property, such as the luminosity, the BH mass, or the quasar's redshift. This would allow for targeted generation of quasar spectra and offers insight into the properties of a particular population. In addition, conditioning the VAE on both redshift and luminosity might mitigate biases inherited from the SDSS selection function and allow the model to learn the relevant quasar physics more easily. We plan to further develop these ideas in the future and implement them in \texttt{QUEST}.
    \item Estimate quasar properties directly from the latent space, as a complementary approach to inferring them to reconstructed spectra. This requires the model to have learnt the relevant physics and probably requires dedicated training datasets. Considering the GP model, for example, the model does not have access to a fully connected parameter space: this is evident from Fig. \ref{fig:MI-z_dist}, where the high-$z$--faint and the low-$z$ regimes are not populated.
\end{itemize}

\section{Conclusion}
\label{sec:conclusion}
In this work, we present a general model for quasar spectra based on an Information Maximising Variational Auto-Encoder architecture. The model is capable of generating realistic quasar spectra that can be post-processed for different purposes, ranging from the generation of synthetic photometry to imputation of BAL features, to the reconstruction of emission line to then estimate the quasar black hole mass. In particular:
\begin{itemize}
    \item we produce three complementary datasets: the General Purpose, the Full Overlap Blue and Full Overlap Red datasets. In all cases, we start from the SDSS DR16Q quasar catalogue, apply quality cuts to select type-1 quasar without absorption systems or intrinsic reddening. The GP dataset is designed to cover the largest wavelength interval, from 980~\AA\ to 5500~\AA\ with the goal of producing a ``jack of all trades'' model. The Full Overlap Blue and Red datasets, instead, are designed to showcase the adaptability of the model and geared towards more specific science cases, namely imputation of BAL features and reconstruction of emission lines with the purpose of estimating the corresponding quasar black hole mass;
    \item after training the model and verifying that it provides an accurate reconstruction of the SDSS spectra, we investigate whether the latent space correlates with physical properties. To do so, we first develop an intuition for which spectral features are affected by each latent dimension, by varying a single latent space dimension, reconstructing the spectra and inspecting the results. We then reduce the dimension of the VAE latent space using UMAP and look for correlation with quasar properties in the resulting embedding. Finally, we apply \texttt{GMM-MI}, an estimator for mutual information, to robustly quantify the correlation between latent space dimensions and quasar properties. Through these tests, we identify correlation between latent dimensions and quasar continuum slope, continuum luminosity, absolute \textit{i}-band magnitude, black hole mass, emission line equivalent width and line luminosity. Although it is possible that the model picked up physical quasar physical properties, we cannot exclude the fact that at least part of these correlations stem from the SDSS selection function. The strong correlation between redshift and UMAP representation (Fig. \ref{fig:UMAP_color_coding_subset}) could hint towards this direction.
    \item to showcase their capabilities, we employ the model trained on the GP dataset to generate synthetic quasar photometry, the one trained on the FOB dataset to input BAL features, and the one trained on the FOR dataset to reconstruct emission lines in order to estimate the black hole masses. We find that the photometry estimated from the quasar spectra faithfully reproduces the SDSS colours of the low-$z$ quasar and the colours of the quasars with $z > 5.3$ from \citet{fan_quasars_2023}. The FOB model, while providing a satisfactory reconstruction in most cases, struggles to accurately reconstruct asymmetric and blue-shifted emission lines (such as the \Civ). The BH masses obtained from fitting FOR spectra are generally in good agreement with the BH masses we estimate from the real SDSS quasar spectra (albeit overestimated by a factor of $\sim 1.25$), with the most significant differences arising for objects with the largest BH masses. Detailed investigation of these problems is beyond the scope of this paper, but it is possible that the lack of training data hampers the capabilities of the model.
\end{itemize}

In the future, we aim to further perfect the model, by expanding the training dataset to include more data, generate targeted datasets (that include, for instance, quasar with broad absorption lines, weak emission lines or that are reddened) and improve the current architecture to include recent advances in machine learning. This will allow, for example, to efficiently select these sources from present and future astronomical surveys.

\section{Acknowledgements}
The code underlying this work makes significant use of the following open-source projects: \texttt{numpy} \citep{harris_array_2020}, \texttt{astropy} \citep{robitaille_astropy_2013, collaboration_astropy_2018, collaboration_astropy_2022}, \texttt{matplotlib} \citep{hunter_matplotlib_2007} and \texttt{pandas} \citep{the_pandas_development_team_pandas-devpandas_2025}.

This work has been supported by the Deutsche Forschungsgemeinschaft (German Research Foundation; Project Nos. 518006966 to J.-T.S. and FG, and 506672582 to S.E.I.B.). LLS acknowledges support by the Deutsche Forschungsgemeinschaft (DFG, German Research Foundation) under Germany’s Excellence Strategy – EXC 2121 ``Quantum Universe'' – 390833306. JFH acknowledges support from the European Research Council (ERC) under the European Union’s Horizon 2020 research and innovation programme (grant agreement No 885301), from the National Science Foundation (NSF) under Grant No. 2307180, and from NASA under the Astrophysics Data Analysis Programme (ADAP, Grant No. 80NSSC21K1568). RAM acknowledges support from the Swiss National Science Foundation (SNSF) through project grant 200020\_207349.

Funding for the Sloan Digital Sky Survey IV has been provided by the Alfred P. Sloan Foundation, the U.S. Department of Energy Office of Science, and the Participating Institutions. SDSS-IV acknowledges support and resources from the Center for High Performance Computing  at the University of Utah. The SDSS website is www.sdss4.org. SDSS-IV is managed by the Astrophysical Research Consortium for the Participating Institutions of the SDSS Collaboration including the Brazilian Participation Group, the Carnegie Institution for Science, Carnegie Mellon University, Center for Astrophysics | Harvard \& Smithsonian, the Chilean Participation Group, the French Participation Group, Instituto de Astrof\'isica de Canarias, The Johns Hopkins University, Kavli Institute for the Physics and Mathematics of the Universe (IPMU) / University of Tokyo, the Korean Participation Group, Lawrence Berkeley National Laboratory, Leibniz Institut f\"ur Astrophysik Potsdam (AIP),  Max-Planck-Institut f\"ur Astronomie (MPIA Heidelberg), Max-Planck-Institut f\"ur Astrophysik (MPA Garching), Max-Planck-Institut f\"ur Extraterrestrische Physik (MPE), National Astronomical Observatories of China, New Mexico State University, New York University, University of Notre Dame, Observat\'ario Nacional / MCTI, The Ohio State University, Pennsylvania State University, Shanghai Astronomical Observatory, United Kingdom Participation Group, Universidad Nacional Aut\'onoma de M\'exico, University of Arizona, University of Colorado Boulder, University of Oxford, University of Portsmouth, University of Utah, University of Virginia, University of Washington, University of Wisconsin, Vanderbilt University, and Yale University.

\bibliographystyle{aa}
\bibliography{general_bib}

\begin{appendix}
\onecolumn

\section{Median composite of the General Purpose training dataset}
Here and online in electronic form, we provide the data underlying Fig. \ref{fig:medianCompositeSpec}. The columns ``Rest-Frame wavelength'' and ``\# of spectra'' contain, respectively, the rest-frame wavelength array we use to sample each spectra and the number of spectrum contributing to the composite. The column ``Flux density'' contains the median flux density, per pixel, of the GP sample. Percentiles columns contain the n$^{\rm th}$ flux percentile, as shown in Fig. \ref{fig:medianCompositeSpec}. The units of the median flux (and of each percentile column) are arbitrary, as all spectra are normalised between 2350\AA and 2360\AA.

\begin{table}[ht]
	\caption{Quantities used to produce Fig. \ref{fig:medianCompositeSpec}}
	\label{tab:medianCompositeSpec}
	\centering
	\begin{tabular}{c c c c c c c}
        \toprule
        Rest-Frame wavelength [\AA] & Flux Density [A.U.] & \# of spectra & 1$^{\rm st}$ perc. & 16$^{\rm th}$ perc. & 84$^{\rm th}$ perc. & 99$^{\rm th}$ perc. \\
		\midrule
        980.000                     & 3.515               & 107          & 1.736        & 2.396             & 4.716             & 7.129             \\
        980.458                     & 3.558               & 111          & 1.743        & 2.442             & 4.761             & 7.171             \\
        980.916                     & 3.571               & 121          & 1.752        & 2.413             & 4.700             & 7.160             \\
        ...                         & ...                 & ...          & ...          & ...               & ...               & ...               \\
		\bottomrule
	\end{tabular}
\end{table}

\section{Sampled median compared to median for FOR and FOB models}
\begin{figure}[ht]
    \centering
    \includegraphics[width=\textwidth]{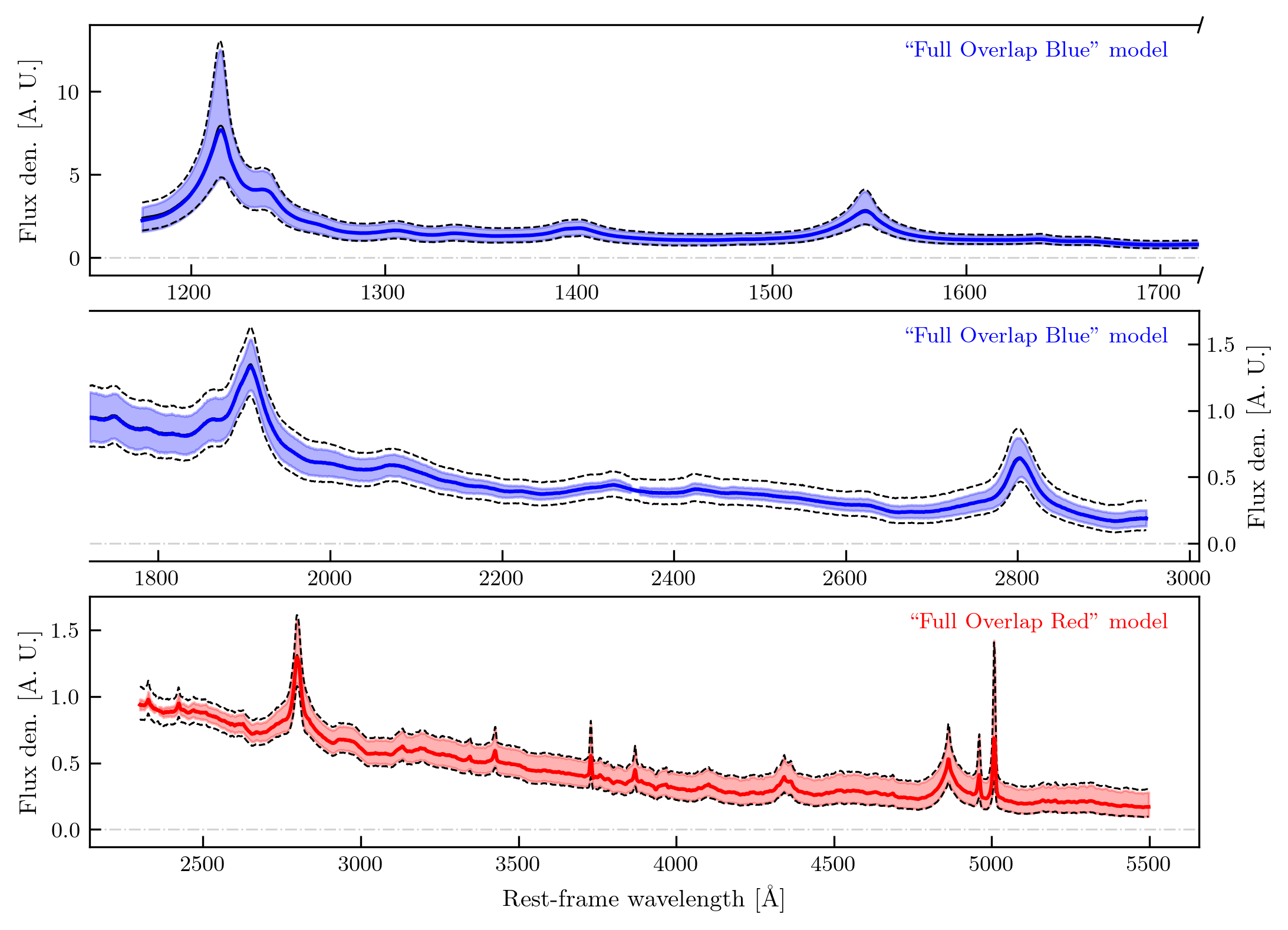}
    \caption{Sampled spectra from the FOR and FOB models compared to the input spectra. In all panels, the black solid and dashed lines indicate the median, 16$^{\rm th}$ and 84$^{\rm th}$ percentile of the input data. The solid grey line represents the median spectrum of 10~000 realisations sampled from the VAE, while the shaded area encompasses the 16$^{\rm th}$ and 84$^{\rm th}$ percentile of the same sampled data.}
    \label{fig:FOB_FOR_sampled}
\end{figure}

In Fig. \ref{fig:FOB_FOR_sampled} we present the comparison between the median of sampled spectra and the median of the input data for the FOR and FOB datasets. As was the case for the GP dataset, the input spectra are normalised in a window between 2350~\AA{} and 2360~\AA{}.

\section{Latent space distributions}
\begin{figure}[ht]
    \centering
    \includegraphics[width=\textwidth]{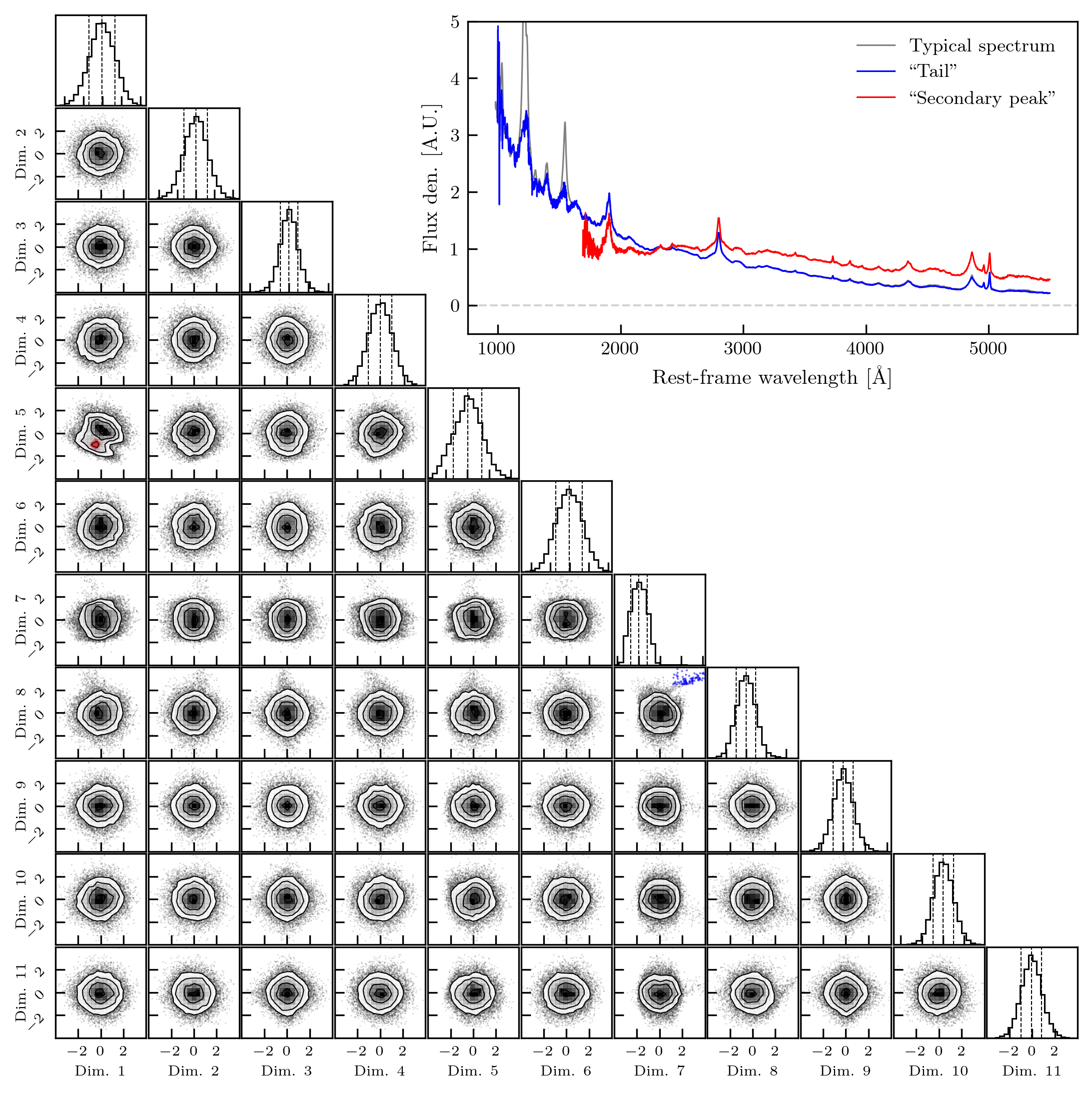}
    \caption{Corner plot showing the latent space dimensions for the GP model. We highlight in red the ``secondary peak'' in LD1--LD5 and in blue the extended tail in LD7. In the top right we show the median SDSS spectra populating the ``tail'' (blue) and the secondary peak (red). These appear to be quasar with either a weak or absorbed \lya{} emission line, and reddened spectra that were not excluded by our preprocessing.}
    \label{fig:latent_space_dims_corner}
\end{figure}

In Fig. \ref{fig:latent_space_dims_corner} we present a corner plot showing each latent space dimension for the GP model. Almost all the latents are approximately Gaussian, with the exception of LD5, featuring an asymmetric distribution (corresponding to the secondary peak highlighted in red) and to a lesser extent LD7, featuring an extended ``tail'', highlighted in blue. We visualise the spectra corresponding to these features by plotting the median spectrum, and find them to correspond to low-$z$, reddened spectra and spectra that have absorption in the \lya{} emission line, respectively.

\section{Latent space variations}
\begin{figure}[ht]
    \centering
    \includegraphics[width=\textwidth]{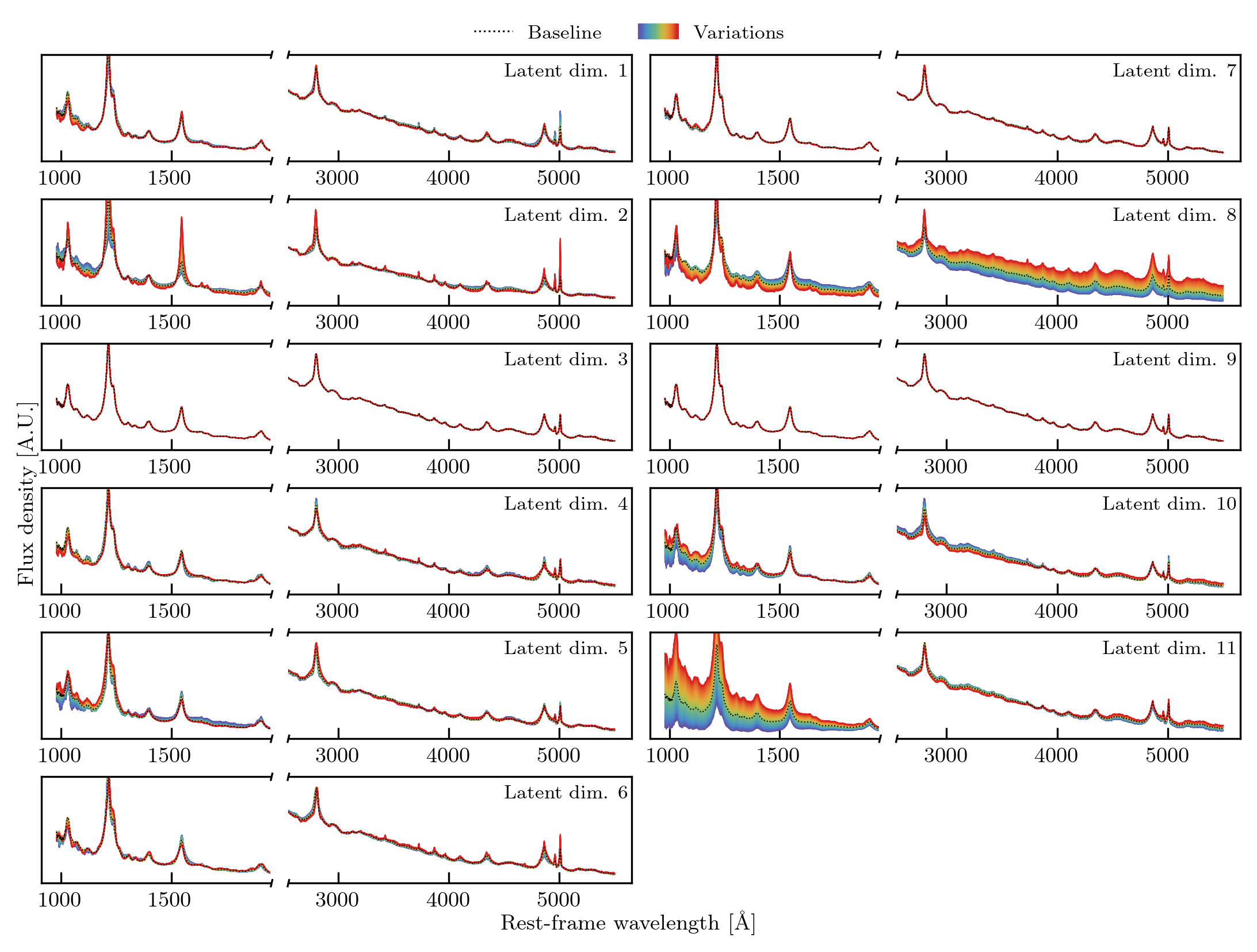}
    \caption{Latent space variations for the eleven dimensions of the GP model.}
    \label{fig:latent_space_variations_all}
\end{figure}

We present in Fig. \ref{fig:latent_space_variations_all} the latent space variations for all the eleven dimension of the GP model. As for Fig. \ref{fig:latent_space_variations}, we compute the variations by decoding mock latent space vectors where only one dimension is varied. The varied dimension is marked in the upper right corner. 

\section{List of SDSS identifiers in each HDBScan cluster}
\begin{table}[ht]
	\caption{SDSS identifier in each cluster. The full table will be made available online.}
	\label{tab:sdss_identifiers_hdbscan}
	\centering
	\begin{tabular}{c c c}
        \toprule
        Red cluster & Green cluster & Blue cluster \\
		\midrule
        SDSS J225515.37+241011.3 & SDSS J011422.47+303719.1 & SDSS J085402.18+274949.3 \\
        SDSS J010728.57+033348.6 & SDSS J112224.74+491624.2 & SDSS J153751.87+531022.2 \\
        SDSS J081815.99+422245.4 & SDSS J102318.17+074419.1 & SDSS J225612.95+234712.0 \\
        ...                      & ...                      & ...                      \\
		\bottomrule
	\end{tabular}
\end{table}

\FloatBarrier

\section{BAL quasars}
\begin{figure}[ht]
    \centering
    \includegraphics[width=\textwidth]{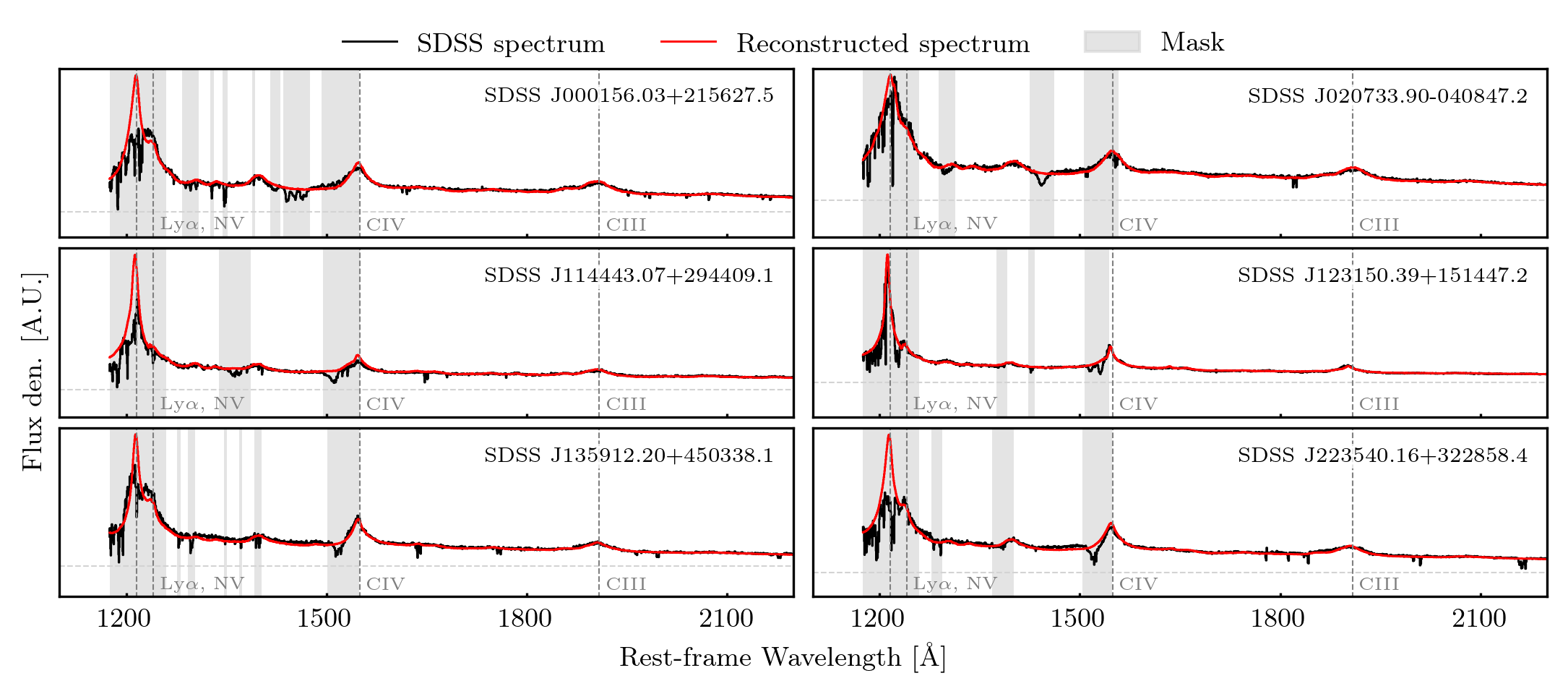}
    \caption{Spectra of the six remaining quasars used to test the imputation capabilities of the model. As in Fig. \ref{fig:BAL_rec_rest}, we show in black the input spectrum, in red the reconstruction, and with the shaded, grey areas the masked regions. The SDSS identifier is indicated in the top right corner.}
    \label{fig:BAL_rec_rest}
\end{figure}

We present in Fig. \ref{fig:BAL_rec_rest} the remaining six spectra of BAL quasars with the reconstruction from the model. In most cases, the model struggles to model the unabsorbed continuum and the emission lines, leading to sub-par reconstructions.

\end{appendix}
\end{document}